\documentclass[11pt,preprint]{aastex}

\usepackage{lscape}
\usepackage{amssymb}
\usepackage[]{hyperref}
\hypersetup{
pdfauthor = {M. Almudena Prieto},
pdftitle = {SED of galaxy cores}
}

\usepackage[all]{hypcap}

\begin{document}
\title{The spectral energy distribution of the central parsecs of the nearest AGN}  
\author{M. A. Prieto (1), J. Reunanen (2), 
 K. R. W. Tristram (3), N. Neumayer (4), J. A. Fernandez-Ontiveros (1) , M. Orienti (1) \& K. Meisenheimer (5)}
 
\affil{(1) IAC, Tenerife; (2) Tuorla Observatory, University of Turku; (3) MPIfR, Bonn; (4) ESO, Garching, (5) MPIA, Heidelberg}
\maketitle

\section{Abstract}

Spectral energy distributions (SEDs) of the central few tens of parsec region  of  some of the nearest, most well studied,  active galactic nuclei (AGN) are presented.
These genuine AGN-core SEDs, mostly from Seyfert galaxies,  are 
 characterised by two main features: an IR bump  with the maximum in the 2 -10 $\mu$m range, and an increasing  X-ray spectrum   with frequency in    the  1 to $\sim$ 200 keV region. These  dominant features are common  to  Seyfert type 1 and 2 objects  alike. In detail, type 1 AGNs    are clearly distinguished from type 2s by their high spatial resolution SEDs: type 2  AGN exhibit a sharp drop shortward of 2 $\mu$m, with  the optical to UV region being fully absorbed;  type  1s  show instead a gentle 2 $\mu$m drop   ensued by a secondary, partially-absorbed optical to UV emission bump. On the assumption that the bulk of optical to UV photons  generated in these AGN  are   reprocessed  by dust and re-emitted in the IR in an isotropic manner, the IR bump luminosity  represents  $\gtrsim 70\%$   of the total energy output   in these objects, the second energetically important contribution  are the high energies   above 20 keV. \\

Galaxies selected by their warm infrared colours, i.e. presenting a
relatively-flat flux distribution in the 12 to 60 $\mu$m range  have often being classified as   active galactic nuclei (AGN). The results from these high spatial resolution SEDs
 question this criterion as a general rule.  It is found that  the intrinsic  shape  of the infrared  spectral energy distribution of an AGN and inferred bolometric  luminosity  largely   depart from those derived  from   large aperture data.  AGN  luminosities can be overestimated by up to two orders of magnitude if relying on  IR satellite data.
We find these differences  to be critical    for AGN luminosities
below or about $10^{44}$ erg~ s$^{-1}$. Above this limit,     AGNs tend to dominate the light of their host galaxy regardless of the  integration aperture size used. Although the number of objects presented in this work is small,  we tentatively mark   this luminosity as a threshold to identify galaxy-light-  vs AGN- dominated objects. \\




\section{Introduction \label{section1}}  

The study of the spectral energy distributions (SED) over the widest possible spectral range is an  optimal way to characterise the properties of galaxies in general. Covering the widest spectral range is the key to  differentiate physical phenomena which  dominate at specific spectral ranges: e.g.  dust emission in the IR, stellar emission in the optical to UV, non-thermal processes in the X-rays and radio,  and to interrelate them as most of these  phenomena  involve radiation  reprocessing from a spectral range into another. The availability of the  SED of a galaxy   allows us to determine basic parameters such as its bolometric luminosity (e.g. Elvis et al. 1994;  Sanders \& Mirabel 1996; Vasudevan \& Fabian 2007), and via    modelling of the SED, its star formation level, mass  and  age (e.g. Bruzual \& Charlot 2003; Rowan-Robinson et al. 2005;  Dale et al. 2007).

The construction of bona-fide SEDs is not easy  as it involves  data acquisition  from different ranges of the electromagnetic spectrum using very different telescope infrastructure. That already introduces a further complication as  the achieved spatial
resolution, and with it the aperture size used,  vary with  the spectral range. SEDs based on the integration of the overall galaxy light may be very different from those extracted from only a specific region, for example the nuclear region. In this specific case,   the aperture size matters a lot, as different light sources, such as circumnuclear star formation, the active nucleus, and the subjacent galaxy light, coexist on small spatial scales and may contribute to the total nuclear output with comparable energies (e.g. Genzel et al.  1998; Reunanen et al. 2009). 

In the specific case  of SEDs of AGN, it is often assumed that the AGN light dominates the integrated light of the galaxy at  almost any spectral range and for almost any aperture. This assumption becomes mandatory  at certain spectral ranges, such as  the high energies, the extreme UV or the mid-to-far-IR, because of  the spatial resolution limitations  imposed by the available instrumentation,  which currently lies in the several arcsecs  to  arcminutes  at these wavelengths. In the mid- to far- IR in particular, the available data, mostly  from IR satellites, are limited to spatial resolutions of a few arcsecs at best. Thus, the associated SEDs include  the contribution of  the host galaxy, star forming regions, dust emission and the AGN, with the first two components being measured over different spatial scales in the galaxy depending on the object distance and the spatial resolution achieved at a given IR wavelength. 

In  spectral ranges where high spatial resolution is readily available,  the importance, if not dominance, of circumnuclear star formation relative to that of the AGN has become  clear  in the UV to optical range (e.g. Munoz-Marin et al. 1997), or in the near-IR  (Genzel et al. 1998). In the radio regime, the comparison of low- and high- spatial-resolution maps shows  the  importance of the diffuse circumnuclear emission and the emission from  the jet components  with respect to that of
  the  core itself (e.g Roy et al., 1994; Elmouttie et al, 1998; Gallimore et al. 2004; Val, Shastri \& Gabuzda, 2004).
Even with  low resolution data a major concern shared by most works is the relevance of the host galaxy contribution to the nuclear integrated emission  from the UV to optical to IR.  To overcome these mixing effects introduced by poor  spatial resolution,  different   strategies or assumptions have been followed  by the community. In quasars, by their own nature,  the dominance of the   AGN light over the integrated galaxy light at almost any wavelength is assumed; conversely, in lower luminosity AGN,   the contribution of different components is assessed via modelling of the integrated light (Edelson \& Malkan 1986;  Ward et al. 1987; Sanders et al. 1988; Elvis et al. 1994; Buchanan et al. 2006 among others). 

In this paper,  we attempt to provide a best estimate of  the AGN light  contribution on very nearby AGN by using    very high  spatial resolution data  over a wide range of the electromagnetic spectrum. Accordingly,  SEDs  of the central few hundred parsec  region of some of the nearest
and brightest AGN  are compiled.   The work is motivated by the current possibility to obtain subarcsec resolution data in the near-to-mid-IR of  bright   AGN, and thus at  comparable resolutions to those available with  radio 
interferometry and  the HST in the optical to UV wavelength range. This is possible thanks to the use of adaptive optics in the near-IR, the diffraction limit resolutions provided by  8 -10 m telescopes in the mid-IR as well as  interferometry in the mid-IR. 

The selection of targets is driven by the requirements imposed by the use of adaptive optics in the near-IR, which limits the observations to the availability of having bright point-like targets with magnitudes   V$~< $15 mag in the field, and the current flux detection limits in  mid-IR ground-based observations. AGN in the near universe are sufficiently bright  to  satisfy those criteria.
The  near- to mid- IR high resolution  data used in this work come mostly from the ESO Very Large Telescope (VLT), hence this study relies on Southern targets, all well known objects, mostly Seyfert galaxies: Centaurus A, NGC 1068, Circinus, NGC
1097, NGC 5506, NGC 7582, NGC 3783, NGC 1566 and NGC 7469. For comparison purposes, the SED of the quasar 3C 273 is also included. 

The compiled SEDs make use of the highest spatial
resolution data available with current instrumentation across the electromagnetic spectrum. The main sources of data include: VLA-A array and ATCA data in radio,
VLT diffraction-limited images and VLTI interferometry in the mid-infrared (mid-IR),   VLT adaptive-optics images in the near-infrared,  and \textit{HST} imaging and spectra in the optical-ultraviolet.
Although  X-rays and $\gamma$-rays do not provide such a fine resolution,  information  when available for these galaxies  are also included in the SEDs on the assumption that above 10 keV or so we are sampling the  AGN core region. Most of the data used comes from the Chandra and INTEGRAL telescopes.
 
The novelty in the analysis  is the spatial resolutions achieved  in the infrared (IR), with  typical  full-width at half-maximum (FWHM) $\lesssim $ 0.2 arcsec in the 1--5 $\mu$m, $< $0.5 arcsec in the 11 -- 20 $\mu$m. 
The availability of IR images  at  these spatial resolutions 
allow us to pinpoint the true spatial  location of the  AGN -- which happens not to have an  optical counterpart in most of the type 2 galaxies studied --  and extract its luminosity within aperture diameters of  a few tens of   parsec.
The new compiled SEDs are presented in sect. 3. Some major differences but also similarities between the SEDs of  type 1 and type 2 AGN  arise at these resolutions.  These are presented and discussed in sections 4 and 5.  
 The    SEDs and the inferred nuclear luminosities  are further compared with those extracted in the mid-to-far IR from large aperture data from IR satellites, and the differences discussed in sect. 6. 

  Throughout this paper, $H_0 = 70$ km s$^{-1}$ Mpc$^{-1}$ is used. The central wavelength of the near-IR broad band filters used are: $I$-band (0.8 $\mu$m), $J$-band (1.26 $\mu$m), $H$-band (1.66 $\mu$m), $K$-band (2.18 $\mu$m), $L$-band (3.80 $\mu$m) and $M$-band (4.78 $\mu$m). \\

\section{The high spatial resolution SED:  data source}

This section describes the data used in building the
high spatial resolution SEDs.  Each AGN is analysed in turn, and the compiled SED is    shown in Fig. 1. The data used in the  SEDs are  listed per each object in tables 3 to 11. 
For each AGN,      an upper limit to  the core    size determined in the near-IR
with adaptive optics and/or in the mid-IR with interferometry is provided first. 
The data sources used in constructing the respective SEDs are described next. When found in the literature,  a brief summary of the nuclear variability levels especially  in the IR  is provided.  This is mostly to asses the robustness of the SED shape and integrated luminosities across the spectrum. Finally, as a by product of the analysis, an estimate of the extinction  in the surrounding of  the nucleus based on near-IR colours derived from the high spatial resolution, mostly $J$--$K$, images  is provided. The colour images used are shown in Fig. 2. These are relative  extinction values, resulting  after comparing the average colour in the immediate surrounding of  the nucleus  with that  at further galactocentric regions, usually within the central few hundred  parsecs. These reference regions are  selected from areas presenting  lower extinction as judged from a visual inspection of the colour images.  The derived extinction does not refer to that in the line of sight of the nucleus, which 
 could be much larger - we do not compare with the nucleus colours  but with those in     its surrounding. 
The simplest approach of considering a foreground dust screen is used. The extinction law presented in Witt et al. (1992), for the UV to the near-IR, and  that in Rieke \& Lebofsky  (1985), for the mid-IR, are used.

For some   objects, the near-IR adaptive optics data used to compile the SED  are presented here for the first time. For  completeness purposes, table 12 is a list of all  the adaptive optics VLT / NACO observations  used  in this work. The list of filters and observations date are provided.  For some objects these data was  already presented in the reference quoted in this table. The nucleus of these AGN was  in all cases bright enough  to be used as a reference  for  the adaptive optics system to correct for the atmospheric turbulence. The optical  nuclear source was used in  all cases with the exception of Centaurus A and  NGC 7582, for which the IR nucleus   was used instead.  The observation procedure and data reduction for the objects presented here for the first time are the same as those discussed in detail in   the references quoted in table 12, we refer  the interested reader to those for further details.


There are five type 2, three type 1 and an intermediate type 1 / LINER AGN  in the sample, in addition to the quasar 3C 273, included  for comparative purposes. The  AGN is unambiguously recognised  in the near-IR images of all objects as the most outstanding source in the field of 26 x 26 $arcsecs^{2}$ covered by the VLT/NACO images. This is especially  the case for the type 2 sources where the AGN  reveals in full realm from 2 $\mu$m longwards. The comparable resolution of the NACO-IR- and the HST-optical- images allows us to search for the optical counterpart of the  IR nucleus, in some cases using accurate astrometry based on other point like sources in the field (e.g Prieto et a; 2004;  Fernandez-Ontiveros et al. 2009 ).  In all the type 2 cases  analysed  the   counterpart optical emission is found to  vanish  shortward of 1 $\mu$m. Accordingly, all the type 2   SEDs show a data gap in the optical - UV region. In addition all the SEDs show a common data gap spanning the extreme UV - soft X-rays region, due to the observational inaccessibility of this spectral range, and in the mid-IR to millimetre range due to the lack of data at subarcsec resolution scales. Surprisingly, for some of these well studied objects, neither   radio data at subarcsec scales exist,  e.g. Circinus galaxy.  In these cases,  to get nevertheless an estimate on the nuclear  power in this range, the available radio data is included in the SED.

The SED are further complemented with available data in the X-rays, which extends up to 100 - 200 keV for all sources. Only for Cen A and 3C 273, the SEDs are further extended to   the gamma-rays, these being the only two so far detected at these high energies. \\



{\it Centaurus A} \\

Centaurus A is the nearest radio galaxy in the Southern 
 hemisphere. The adopted scale is 1 arcsec $\sim$ 16 pc (D = 3.4 Mpc; Ferrarese et al. 2007).  The nucleus of this galaxy begins to show up in the optical longward of 0.7 $\mu$m (Marconi et al. 2000) and so far remains unresolved down to a size $<<$1 pc FWHM  in the 1 --  10 $\mu$m range (from VLT adaptive optics  near-IR imaging by  Haering-Neumayer et al. 2006 and  VLT interferometry in the mid-IR by Meisenheimer et al. 2007).

The currently compiled SED of Cen A  core is shown in Fig. 1. The radio data  are  from VLBA quasi simultaneous observations collected in 1997
 at 22.2 and  8.4 GHz (March epoch was selected) and in 1999 at 8.4, 5 and 2.2 GHz, in all cases  with resolution of a few  milliarcsec (Tingay et al. 2001 and  Tingay
 and Murphy, 2001). These spatial resolutions allow for a better separation of  the core emission from the jet. The  peak values reported by the authors  are   included in the SED.

 In the 
 millimetre range, peak values  from  SCUBA in the 350--850 $\mu$m range from Leeuw's et al. (2002) are  included in the SED. SCUBA has poor spatial resolution,  FWHM $\sim$ 10--15
arcsec,  however we take these measurements as genuine core  fluxes as they follow fairly well the trend defined by  the higher resolution data  at both radio and mid-IR wavelengths. This common trend  is a strong indication that    the AGN is  the dominant light source  in the millimetre range.

In the mid-IR range,   VLT / VISIR diffraction-limited data,  taken on March 2006, at 11.9 and 18.7 $\mu$m, from Reunanen et al. (2009)   are included in the SED.
Further mid-IR data are    from VLTI / MIDI interferometric observations in the 8 -- 12 $\mu$m range taken on February  and May 2005 with resolutions of 30 mas (Meisenheimer et al. 2007). The visibility's analysis indicates that at least 60\%  of the emission detected by MIDI   comes from an unresolved nucleus with a size  at 10 $\mu$m of FWHM $<1$ pc. This is also confirmed by more more recent MIDI observations by Burtscher et al (2009). The SED includes  two sets of MIDI fluxes   measured directly in the  average  spectrum from both periods: 1) the total integrated  flux  in the  MIDI  aperture 
(0.52 $\times$ 0.62 arcsec);  2) the core fluxes measured on  the 
correlated spectrum. At 11.9 $\mu$m, the MIDI total flux
and the VISIR nuclear flux, both from comparable aperture sizes, differ by 15\%. The difference  is still compatible with the photometry errors which in particular may be large in the MIDI spectrum-photometry.

The near-IR is covered with VLT/ NACO adaptive optics data taken in $J$-, $H$-, $K$- and $L$-bands, and in the   narrow-band line-free filter centred at 2.02 $\mu$m.
These are complemented with \textit{HST}/WFPC2
data in the $I$-band (Marconi et al. 2000). Shortward of this
wavelength, Cen A's nucleus is unseen. An upper limit derived from the HST / WFPC2 image at 0.5 $\mu$m is included in the SED.

At high energies,  Cen A's nucleus becomes visible
 again,  as well as its jet.  The SED includes the 1 KeV nuclear flux  extracted  by Evans et al. (2004) from
 \textit{Chandra}  observations in 2001  and and average of the  100 keV fluxes derived  by Rothschild et al. (2006) from  \textit{INTEGRAL} observations   collected   in  2003--2004.
In the gamma-rays, the SED includes \textit{COMPTEL} measurements  in the 1 -- 30 MeV
range taken during the 1991--1995 campaign by Steinle et al. (1998).

For comparative purposes, Fig. 1 includes large aperture data -identified with crosses- in 
 the mid to far IR region selected from \textit{ISO} (Stickel et al. 2004) and \textit{IRAS} (Sanders et al. 2003). For consistency purposes, given the poor spatial resolution of the SCUBA  data, these are also labelled with crosses in the SED.
 
 Cen A is a strongly variable source at high energies, where flux variations could be up to an order of magnitude (Bond et al. 1996). However, during the \textit{COMPTEL} observations used in the SED the observed  variability on
the scales of few days is reported to be at the 2 sigma level at most (Steinle et al. 1998).
Combining OSSE, COMPTEL and EGRET, the gamma-ray luminosity (50 keV  - 1 GeV)  varies by 40\%.
No report on variability monitoring  of this source  in the IR was found. The comparison between the VLT / VISIR data used in this work  and equivalent  mid-IR data obtained four years apart, 2002,   by
Whysong \& Antonucci ( 2004) using Keck 
and  Siebenmorgen et al. (2004) using ESO / TIMMI2,  
indicates   a  difference with VISIR   in the   40 to 50\% range. 
 Cen A's nucleus is the single source in this study which  high spatial resolution  SED, from radio to millimetre to IR, can be fit  with a single synchrotron model (Prieto et al. 2007; see also end of sect. 7 ). Thus a flux difference of a factor of 2 in the IR is consistent  with  genuine nuclear variability  and the dominance of a  non-thermal component in the IR spectral region   of this nucleus.
 

Relative extinction values in the nuclear region of  Cen A were measured from
 VLT / NACO $J$--$K$ (Fig. 2) and $H$--$K$ colour maps. The average reddest colours around the nucleus are $J$--$K$ =  1.5, $J$--$H$ = 1.5 and the bluest colours  
  within $\sim$ 100 pc
distance from the nucleus, are $J$--$K$ = 0.68 and
$J$--$H$ = 1.04. Taking these as reference colours  for Cen A stellar population,   the inferred   extinction  around the nucleus is $A_V\sim$ 5 -- 7 mag. 
By comparison, the  extinction inferred from the depth of the silicate 9.6 $\mu$m  feature  in the line of sight of the nucleus -- from the
VLT / MIDI correlated spectra -- is  a factor 2 to 3  larger, $A_v\sim$ 13 - 19 mag (see Table 2). \\

\begin{figure}[p] 
   \centering
  \plottwo{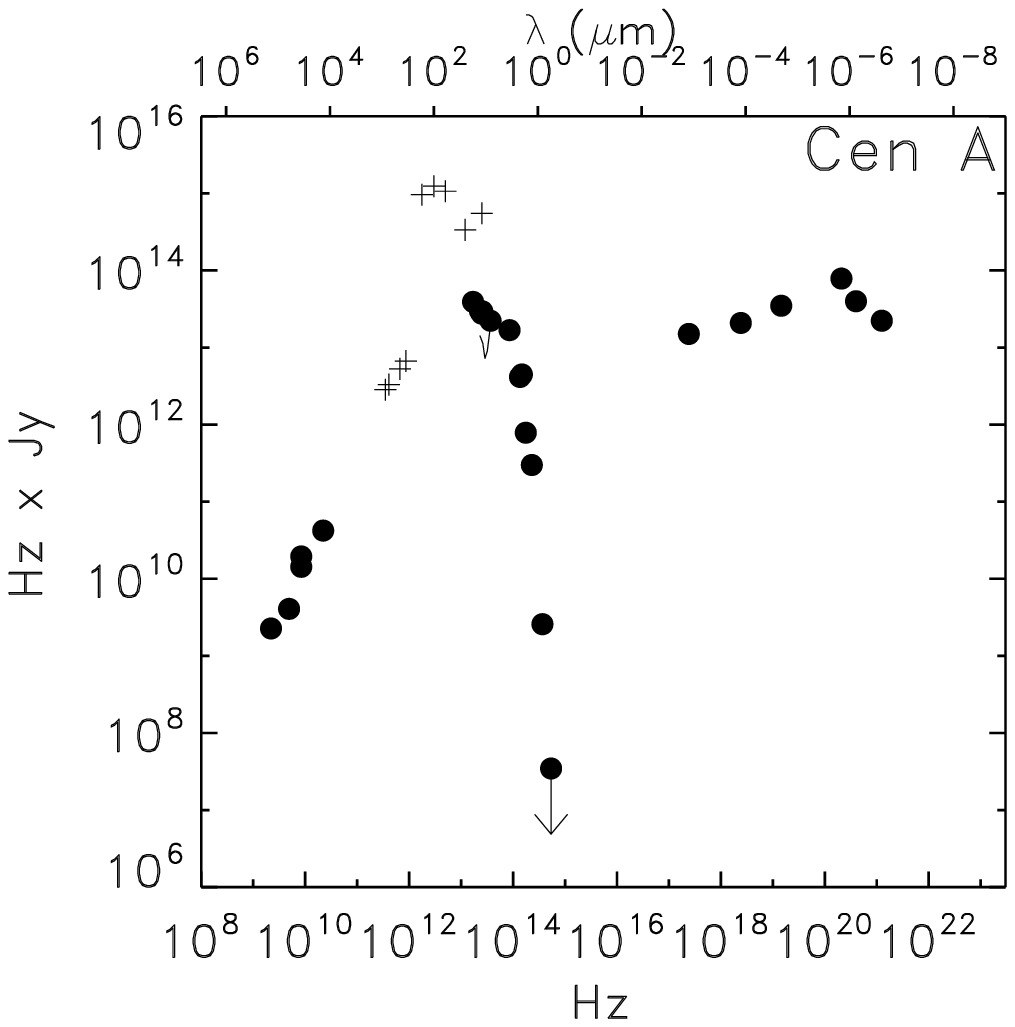}{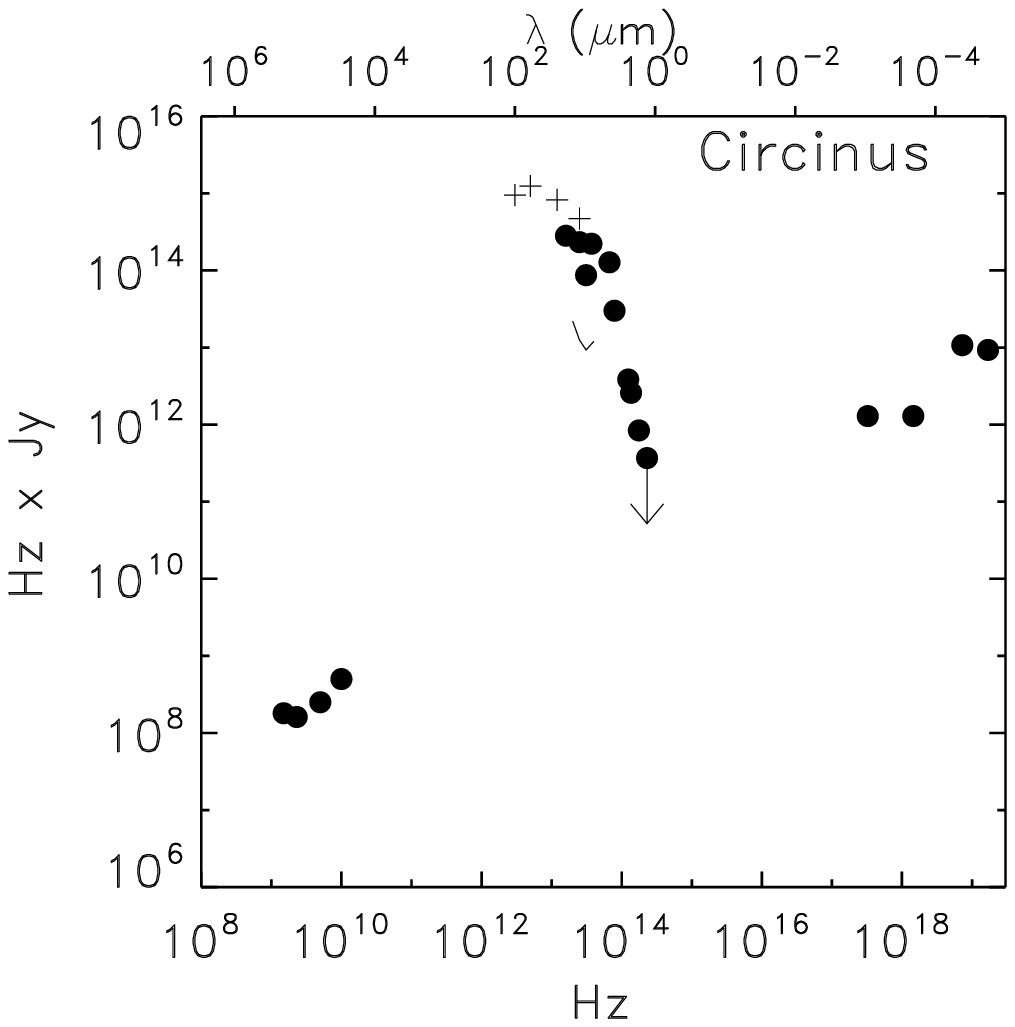}
  \plottwo{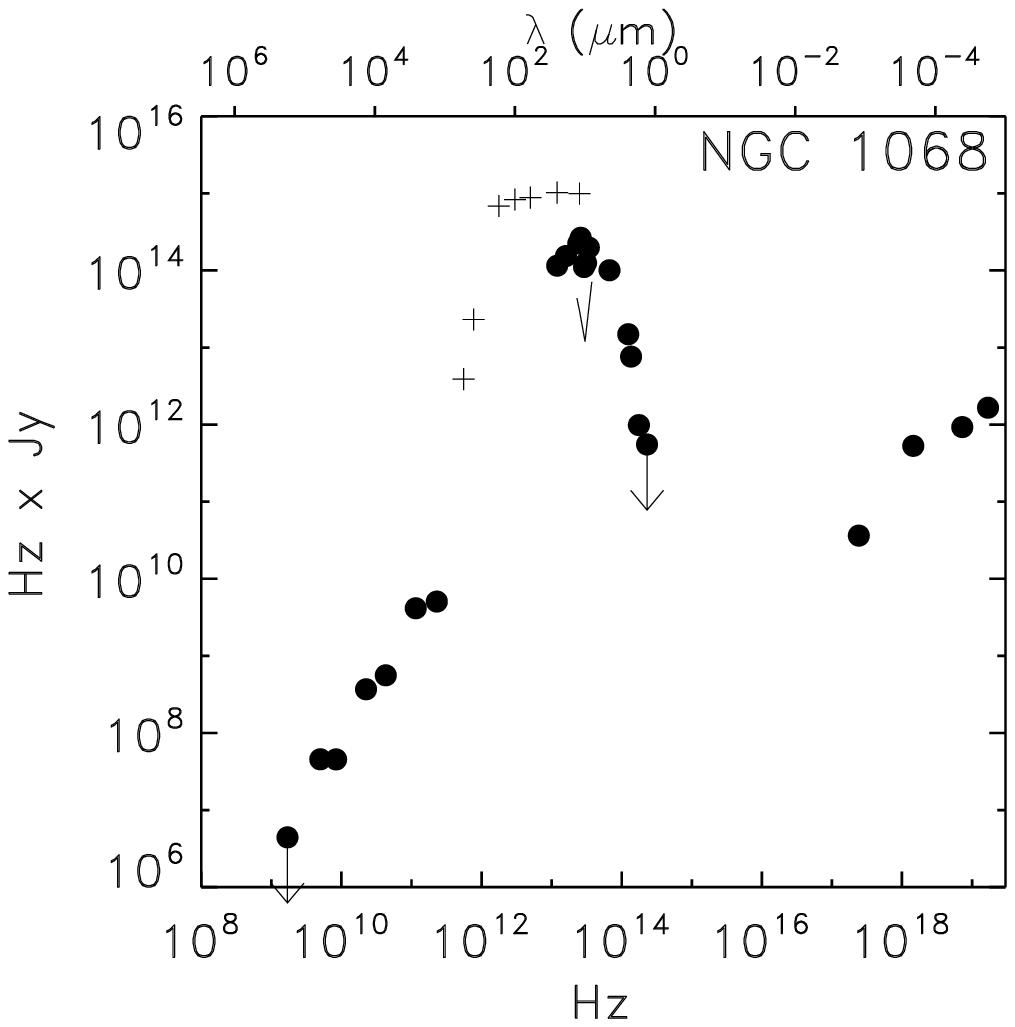}{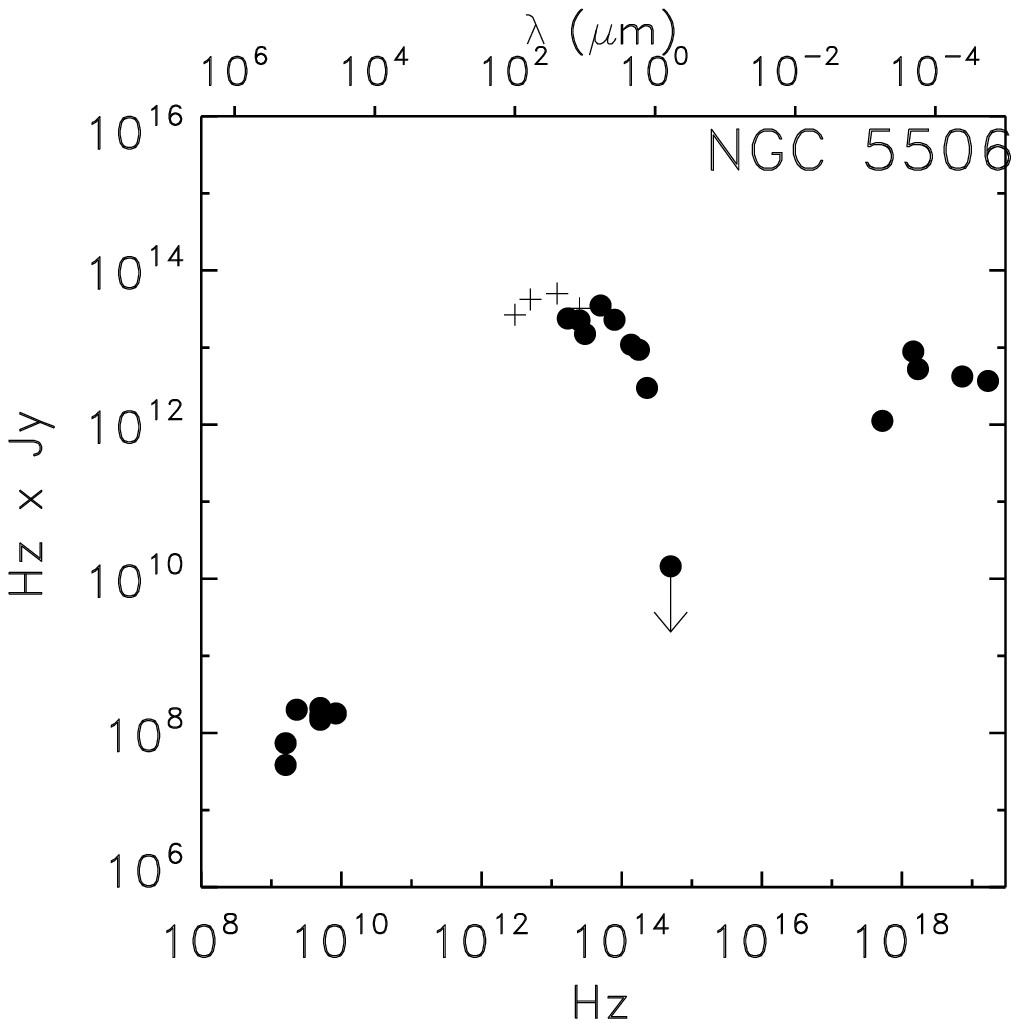}
 \plottwo{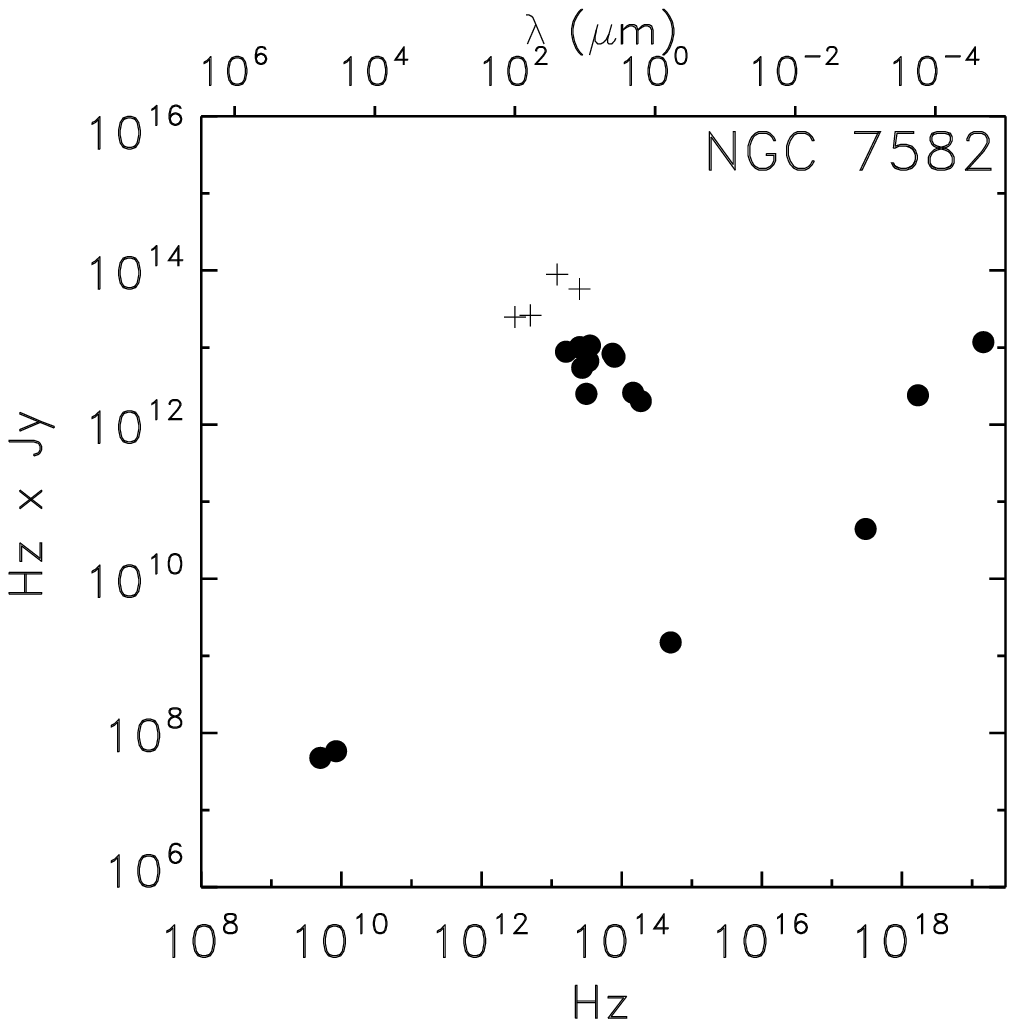}{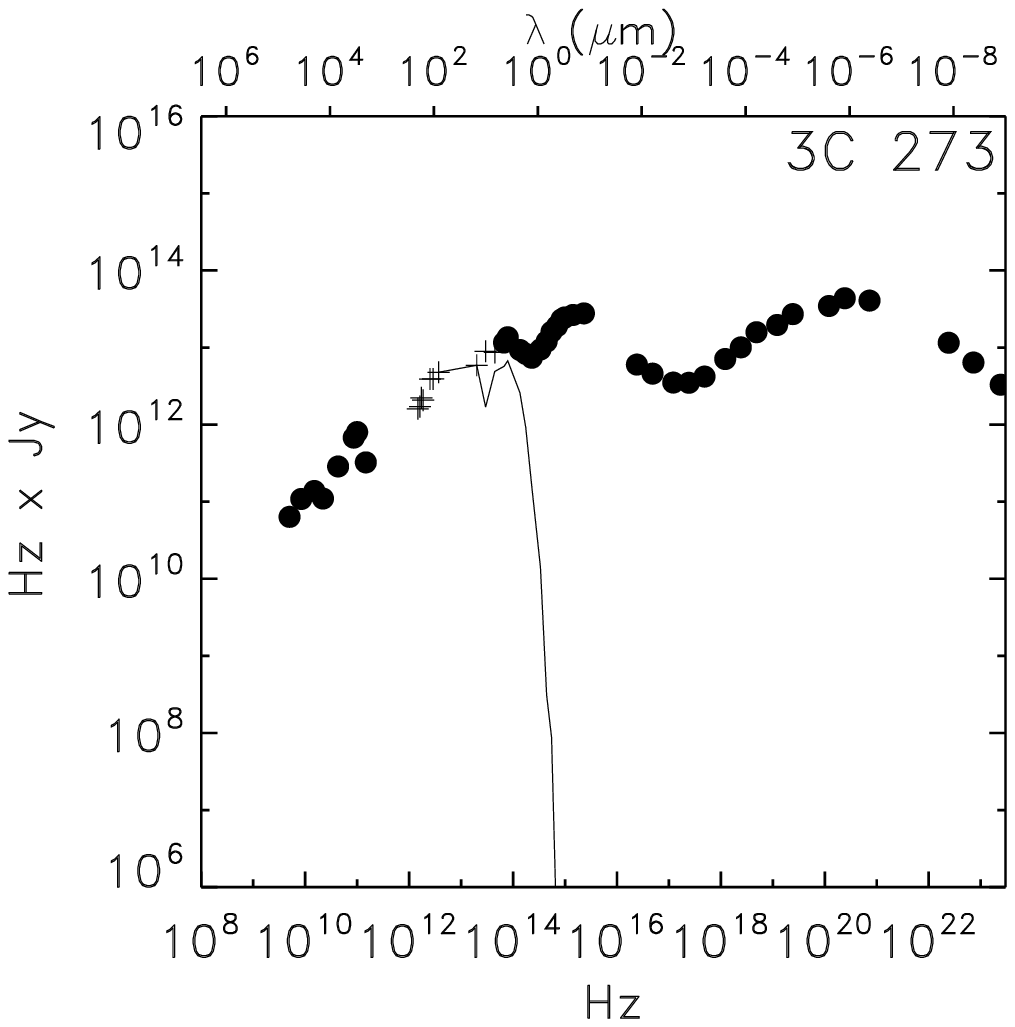}
   \caption{Fig. 1: SEDs of the central parsec-scaled region of  the  AGN in this study: filled points represent the highest 
  spatial resolution data available for these nuclei; the thin  V-shape
  line in the mid-IR region shown in some SEDs corresponds to the spectrum of an unresolved  source as measured by VLTI / MIDI (correlated flux); 
Crosses refer to large aperture data  in the mid-IR (mostly from \textit{IRAS} and \textit{ISO}), and the  millimetre when available. The frequency range is the same in all plots except for Cen A and 3C 273 which extends up to the gamma-rays.  The continuous  line in 3C 273 plot is the SED of this object after applying an extinction Av = 15 mag.}

   \label{Fig. 1}
\end{figure}

\begin{figure}[p] 
\figurenum{1}
   \centering
  \plottwo{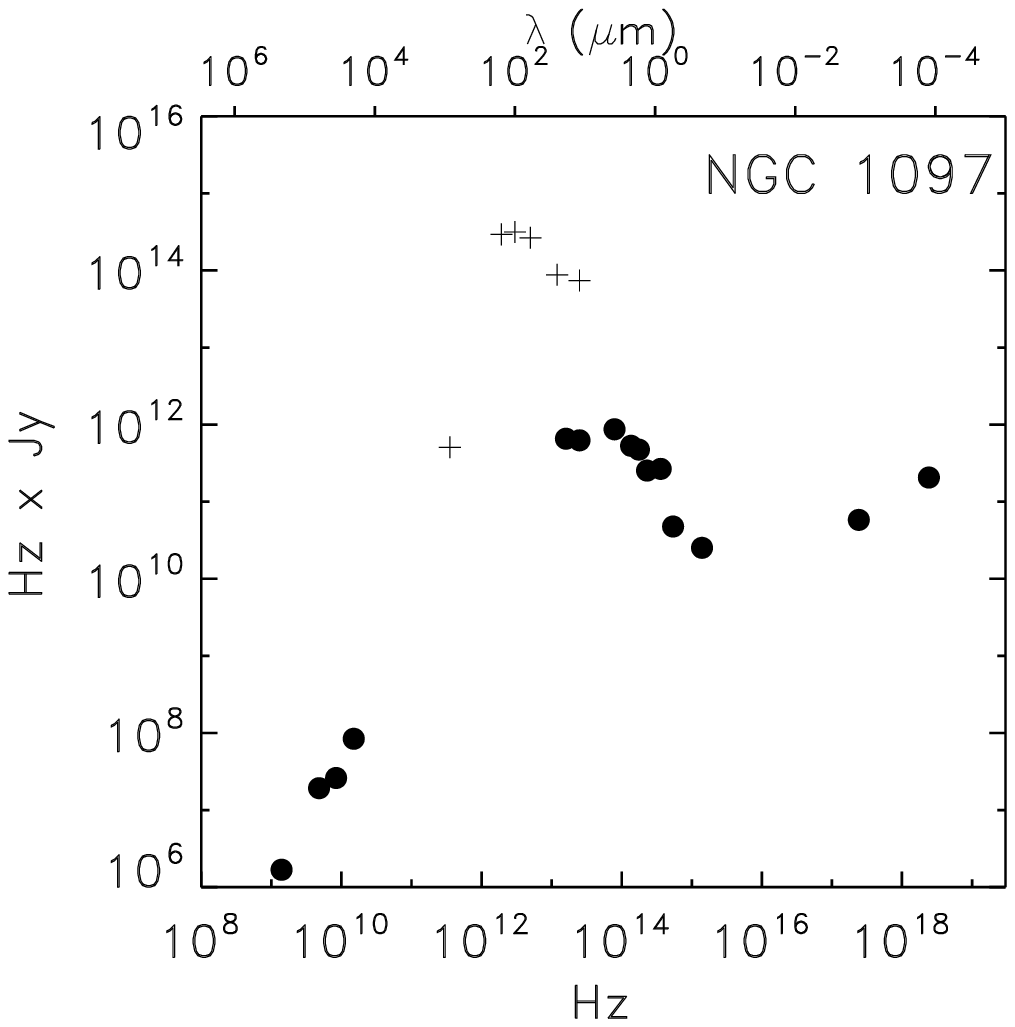}{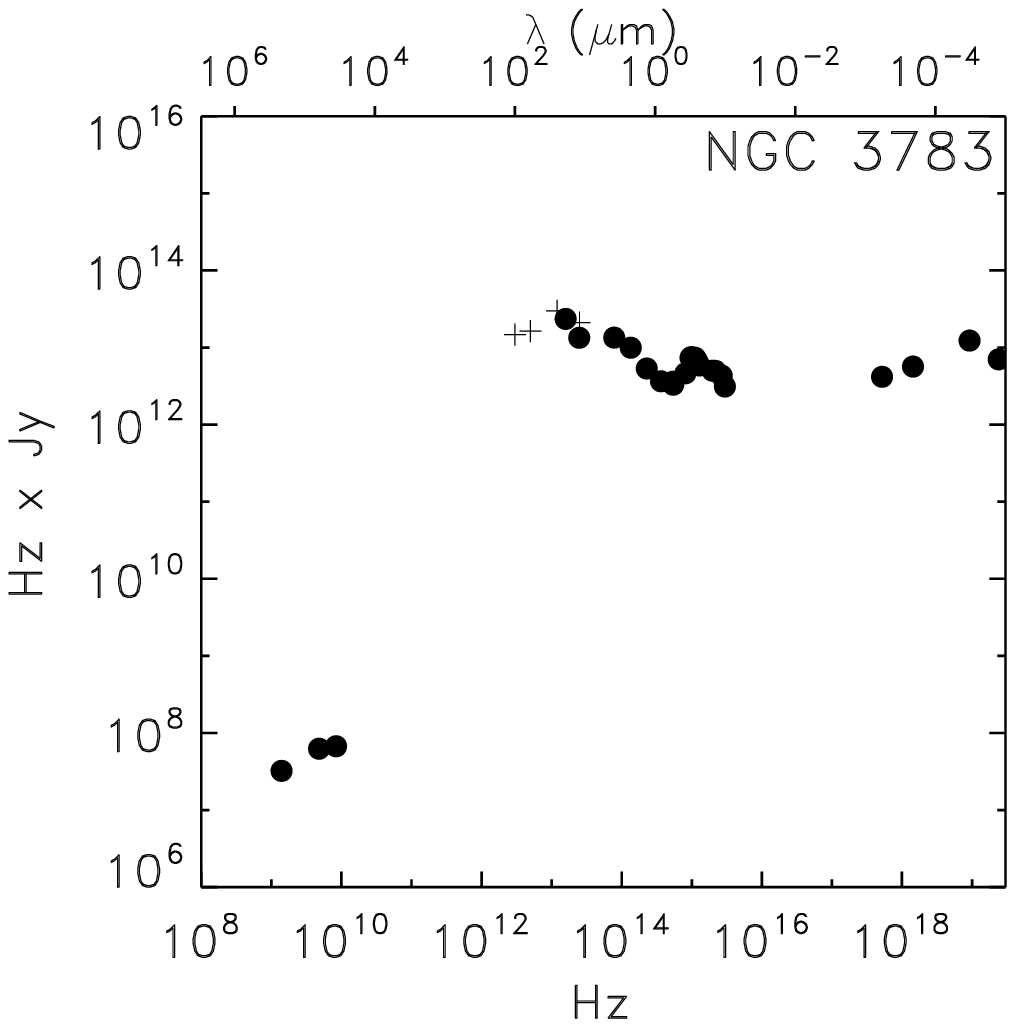}
   \plottwo{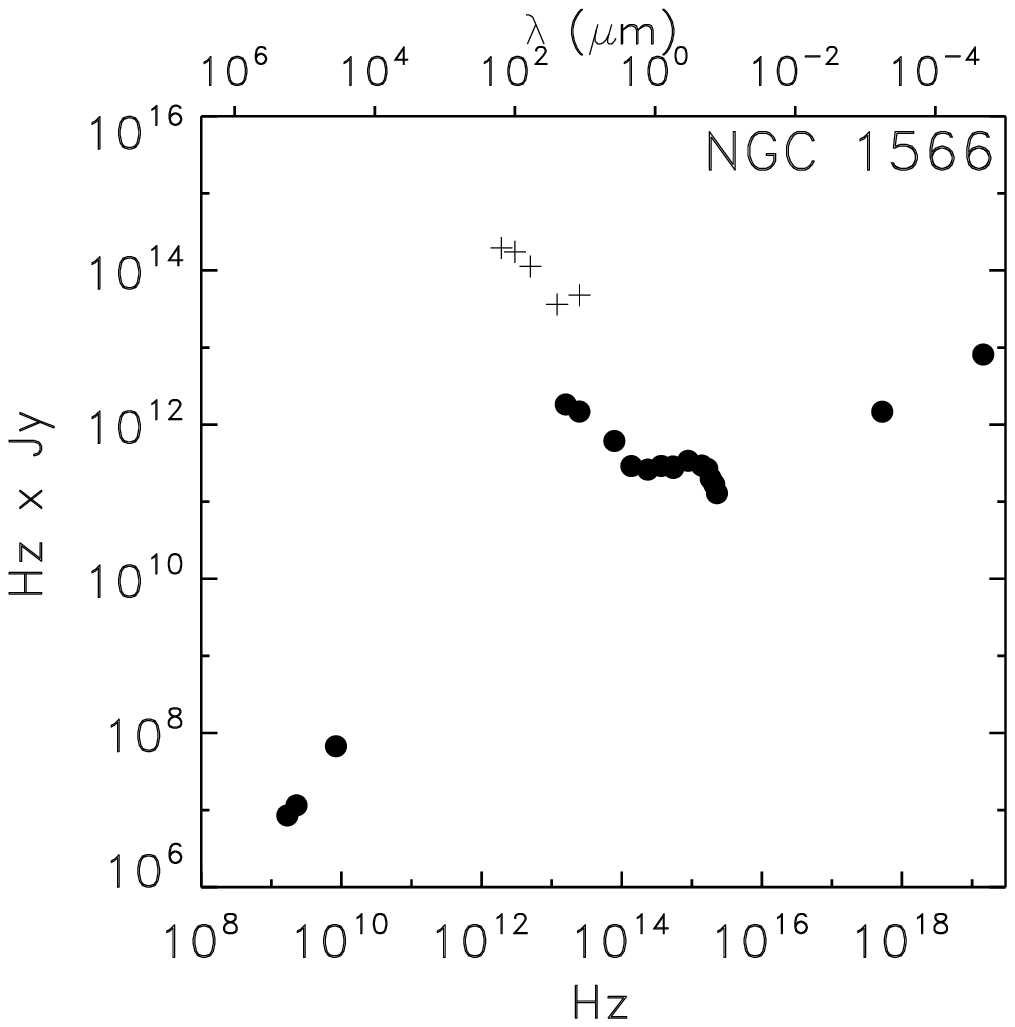}{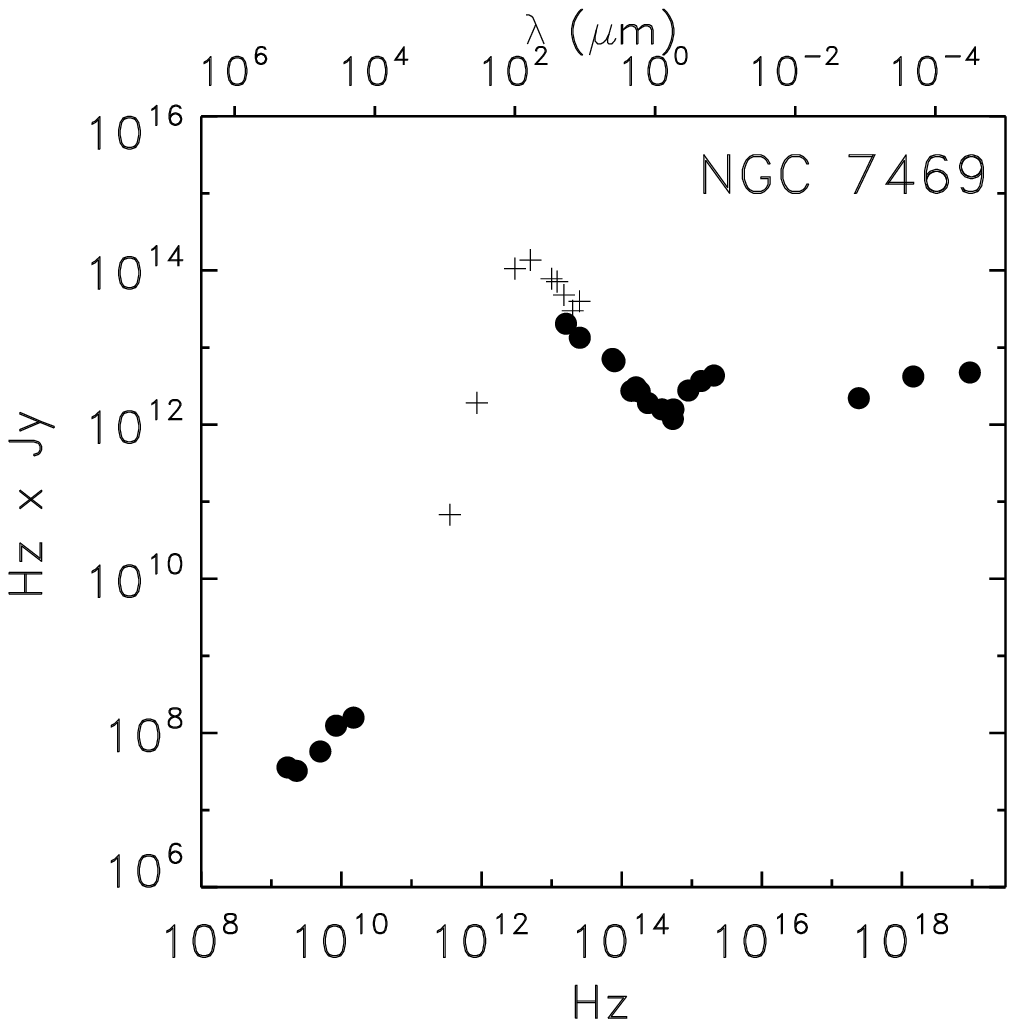}
   \caption{Continued}
   \label{Fig. 1 cont}
\end{figure}
  

\begin{figure}[p] 
   \centering
 \plotone{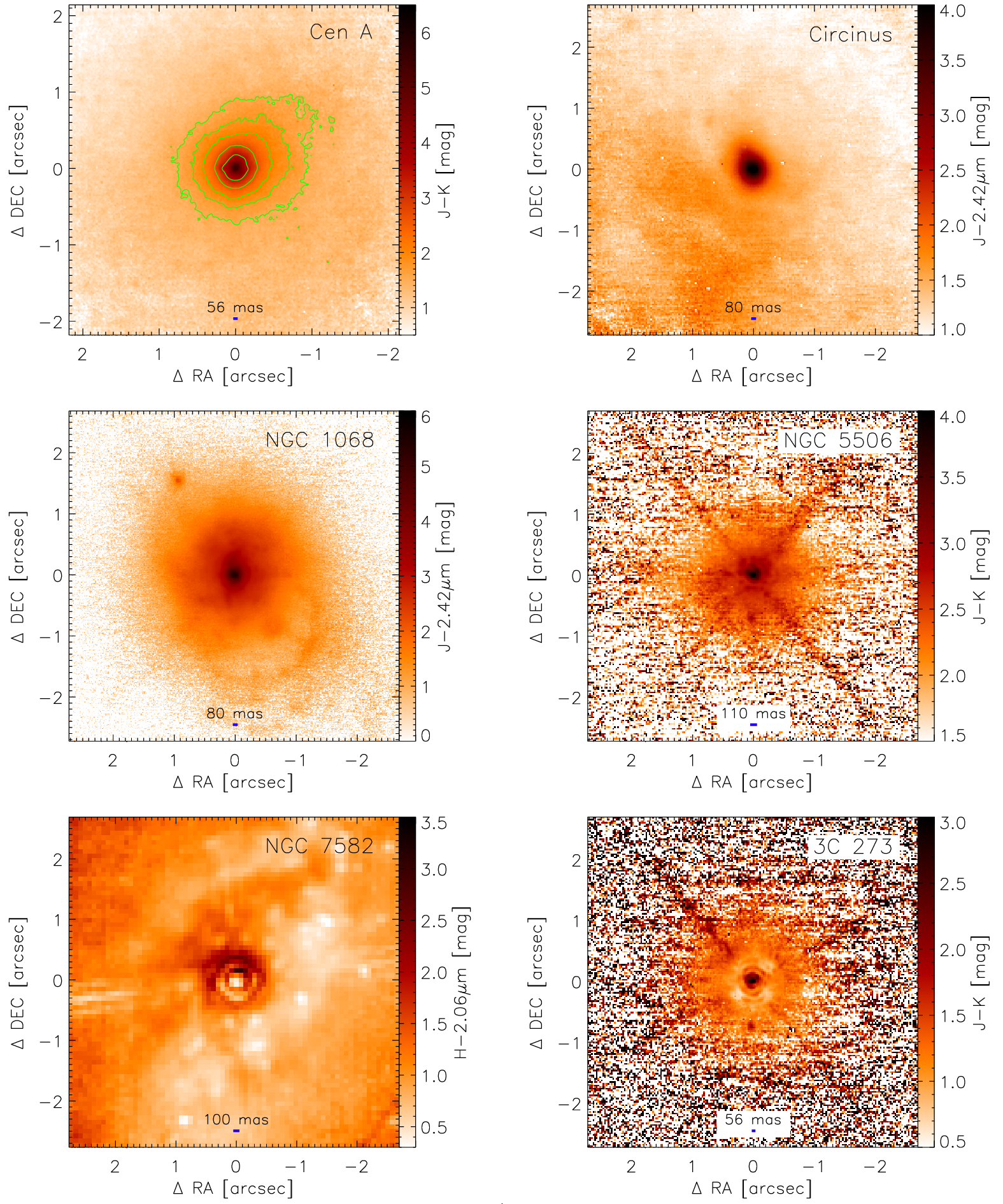}
  \caption{Near-IR colour maps of   the galaxies in this study. Most are   $J$ -- $K$, or $J$ -- narrow band $K$  maps when available. Only the very central 2 to 3 arcsec radius FoV is shown. The small bar at the bottom of each panel indicates the  spatial resolution of the maps, and in turn the upper limit size to the AGN core in the $K$-band.   }
   \label{Fig. 2}
\end{figure}
  
\begin{figure}[p] 
	\figurenum{2}
  \centering
 \plotone{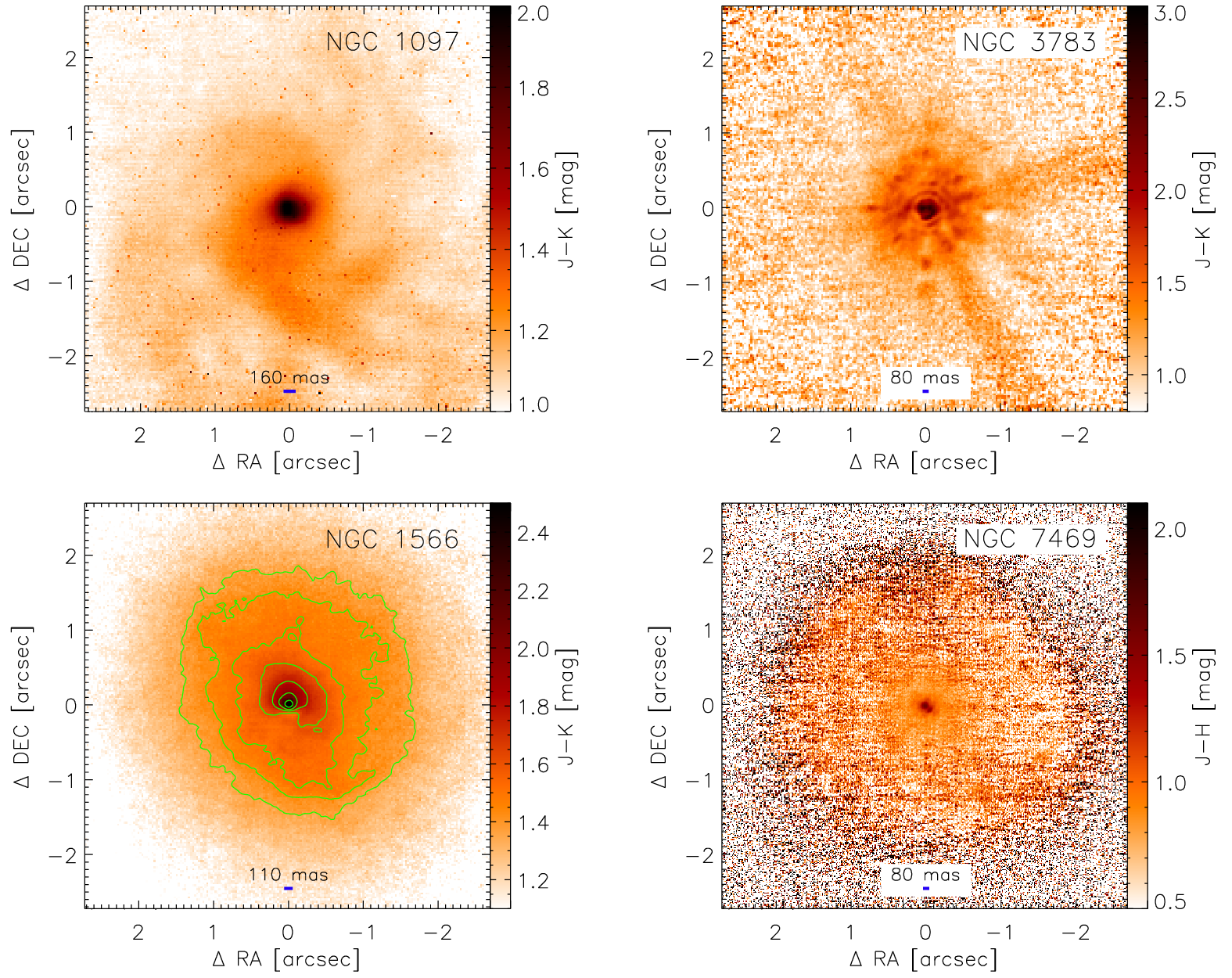}
  \caption{Continued}
   \label{Fig. 2 cont}
\end{figure}

{\it Circinus}\\

Circinus is the second closest AGN in the Southern
Hemisphere. The adopted scale is 1 arcsec $\sim$ 19 pc (D = 4.2 Mpc, Freeman et al. 1977). 
Circinus' nucleus is the only one in our sample  that is spatially resolved in the VLT/ NACO adaptive optics images from 2 $\mu$m onward.  The nucleus resolves into an elongated,   $\sim$ 2  pc size, structure oriented perpendicular to the ionisation-gas cone axis (Prieto et al. 2004). Its  spectral energy distribution is compatible with dust emission at a characteristic temperature of 300 K; this structure has  thus most of the characteristics of the  putative nuclear torus (Prieto et al. 2004). Further VLTI / MIDI   interferometry in the 8 -- 12 $\mu$m range     also resolves the central emission into a structure of the same characteristics (Tristram et al. 2007). 

Circinus' SED is shown in Fig. 1.   
The highest spatial resolution radio data available for this source -- in the range of a few arcsecs only --  is presented by  Elmouttie et al. (1998).  Because of this poor resolution,  only the central peak values   given by that  reference are used in the SED. The  beam sizes at the available  frequencies  are  6 $\times$ 5.4 arcsec at 20 cm, 3.4 $\times$ 3.1 arcsec at 13 cm,  1.4 $\times$ 1.3  at 6 cm, and 0.9 $\times$ 0.8 arcsec at 3 cm.

Mid-IR nuclear fluxes are taken from  VLT/VISIR diffraction-limited images at 11.9 and 18.7 $\mu$m (Reunanen et al. 2009).
Further mid-IR data  are taken  from VLTI/MIDI interferometry spectra covering  the 8 -13 $\mu$m range (Tristram et al. 2007). Two sets of MIDI data are included in the SED:  the correlated fluxes, 
which correspond to an unresolved source detected with visibility of 10\%, and the total fluxes that are measured  within  the MIDI
0.52 $\times$ 0.62 arcsec slit. In the latter case, only  fluxes at  8 and 9.6 $\mu$m are included. For sake of simplicity,  flux measurements at the longer MIDI wavelengths are    not included  as they are in very good agreement with the  VISIR flux at 11.9 $\mu$m  within   5 \%.
Near-IR nuclear fluxes, in the 1 -- 5 $\mu$m range, are taken from   VLT / NACO adaptive optics images by Prieto et al. (2004).
Shortward of 1 $\mu$m, Circinus' nucleus is  undetected, an upper limit at 1 $\mu$m derived from NACO $J$-band image is included in the SED.
 


At the high energies, the SED includes the \textit{ROSAT} 1 keV flux after correction for the hydrogen column density, derived by Contini et al. (1998), and \textit{INTEGRAL} 
 fluxes at different energy bands in the  2 -- 100 keV range (Beckmann et
al. 2006).

For comparative purposes, the SED includes large aperture data  in 
 the mid to far IR region selected from  \textit{IRAS} (Moshir et al. 1990).
 
There are no reports  on significant   variability in Circinus at near-IR wavelengths. In comparing the data analysed in this work, we find that the 12 $\mu$m  nuclear flux measured in 2002 by Siebenmorgen et al. (2004) and  that of VLT/ VISIR collected four years apart, differ by   less 
than 5\%. At the high energies, in
the 20 -- 40 keV range, Soldi et al  (2005) report a  10\% maximum 
variability in a time interval of 2 days.

An estimate of the  extinction toward the nucleus is derived  from  the VLT/NACO $J$--2.42 $\mu$m colour map (Fig. 2 ). In the  surroundings of the nucleus  $J$--2.42$\mu$m  $ ~>$ 2.2 is found. The colours become progressively bluer with increasing distance and  in   relatively clean patches in the central 150 pc radius the average  value is  $J$--2.42 $\mu$m  $\sim$ 1. The  colours comparison yields a relative extinction  of $A_V \sim $ 6 mag, assuming a foreground dust screen. Because of the finest scales, a few parsecs, at which Circinus's nucleus is studied, Prieto et al. (2004)  further considered a simple configuration of the dust being mixed with the stars, in which case, an extinction of   $A_V \sim $ 21 mag is found.   This value is more in line with the range of extinctions     derived from the  optical depth of the 9.6 $\mu$m silicate feature  measured in the VLTI / MIDI correlated spectrum,  $22 < A_V < 33 mag $ (Tristram et al. 2007). \\



{\it NGC 1068}\\

NGC 1068, Circinus and Cen A are the brightest Seyfert nuclei in the sample but NGC 1068 is a  factor four further away,  at a distance of 14.4 Mpc (Bland-Hawthorn et al. 1997). The  adopted scale is 1 arcsec $\sim$ 70 pc.  

NGC 1068's radio  core -- identified with the radio source  S1 -- is resolved     at 5 and
8.4 GHz into an elongated, $\sim$ 0.8 pc disk-like structure, oriented perpendicular to the radio jet axis (Gallimore, Baum \& O'Dea 2004). VLTI / MIDI
interferometry in the  8 -- 12  $\mu$m range resolves the nucleus  into two components: an inner component  of about 1 pc size  and   a cooler, 300 K   disk-like component, with size of 3 x 4 pc (Raban et al. 2009).
At  2$\mu$m speckle observations by Weigelt et al. (2004) resolve the central emission into a compact core with size 
$1.3 \times 2.8 pc$, and  a north- western and south-eastern extensions.
Shortward of 1$\mu$m, and on the basis of the SED discussed below, NGC 1068's  nucleus is fully obscured.

 
The  highest spatial resolution SED of NGC 1068' nucleus from radio to optical is  presented  in Hoenig, Prieto \& Beckert  (2008).   The authors provide  a fair account of the  continuum spectrum from radio to IR to optical SED, by including in their model the  major physical   processes  that must contribute to the integrated core emission:   synchrotron, free-free emission from ionised gas  and  the dust-torus emission. 
The SED included  here is that  presented in Hoenig et al. but further complemented with high energy data. Furthermore, the IR nuclear photometry is  extracted in a different way:   the nucleus of NGC 1068 in the mid- to near- IR  presents  additional emission structure extending, particularly,   North and South of it (Bock et al. 2000; Rouan et al. 2004). To minimise the  contamination by this extended emission, the near-IR nuclear fluxes  are here extracted within an   aperture diameter  comparable with the nuclear FWHM measured at the corresponding wavelength.  

The SED is shown in Fig. 1. The radio data are taken  from
VLA observations at 43 GHz with 50 mas beam (Cotton et al. 2008) and 22 GHz with   75 mas beam  from  Gallimore et al. (1996),   VLBA observations at 8.4 and 5  GHz  from  Gallimore et al. (2004), and VLBA at 1.7 GHz from  Roy et al. (1998).
1 and 3 mm core fluxes  are taken from  Krips et al. (2006), and correspond to  beam sizes larger than 1 arcsec, thus the may include the contribution from the jet.   We still include these data  in the SED as the jet components have a steep spectrum (Gallimore et al. 2004), and thus  their contribution to the core emission is expected to  decrease at these  frequencies.

In the mid-IR, the nuclear fluxes in the  8 to  18  $\mu$m range  are   from Subaru (Tomono et al, 2001), complemented with an additional nuclear flux  at 25  $\mu$m  from Keck (Bock et al. 2000). All these measurements are extracted from deconvolved images in aperture diameters  of $\sim 200$ mas. Despite the uncertainties inherited to the deconvolution process, we note
 that the  fluxes derived by both group of authors at common wavelengths, each using different instrumentation and telescope, differ by $< 30\%$.
The SED  also includes the 8 -- 12 $\mu$m VLTI/MIDI correlated spectrum of the unresolved component sampled at four wavelengths. This correlated spectrum was  derived from a 78 m baseline which provides a spatial resolution of about 30 mas (Jaffe et al. 2004). The total flux measured in the MIDI 500 mas  slit render fluxes larger by  a factor of 2 than those measured by Bock et al. and Tomono et al. in their 200 mas aperture.  This difference is due to the inclusion in the MIDI slit of the light  contribution from   the nuclear extended emission  of  NGC 1068  mentioned above. Thus,  the MIDI total fluxes are not  included in the SED.
Near-IR data, in the 1 -- 5 $\mu$m range,  are from VLT / NACO adaptive optics images at $J$-, $H$-, $K$-, $M$- bands and the  2.42 $\mu$m narrow band line-free filter.  The nucleus is unresolved at the achieved resolutions,  the nuclear fluxes were extracted in aperture diameter comparable to the  nucleus FWHM:  0.1 arcsec in $J$-, $K$-  and 2.42 $\mu$m bands, 0.22 arcsec in $H$-band, 0.16 arcsec in $M$-band. The inspection of the VLT NACO J-band image indicates a relatively  faint source at the position of the K-band nucleus, and thus we take the extracted J-band flux  as an upper limit.  


In the X-ray,  the following measurements are included in the SED: \textit{Chandra} 1 keV flux  from   Young, Wilson \& Shopbell (2001),  \textit{EXOSAT} flux in the 2 -- 10 keV range by Turner \& Pounds (1989),  and  \textit{INTEGRAL} data in the 20 -- 40 keV and 40 -- 100 keV bands from  Beckmann et al. (2006).

The SED includes    large aperture data in the mid- to far- IR   from   \textit{IRAS} (Sanders et al. 2003) and \textit{ISO} (Stickel et al. 2004) and  in the sub-millimetre     from Hildebrand et al. (1977).

Monitoring of the source in the near-IR by Glass (2004) indicates long-term variability, with an increase by a factor of two in flux in the period 1970 - 1995.\\ 
Fig. 2 shows the   VLT / NACO $J$ -- 2.42-$\mu$m colour map. Contrary to other AGN in the  sample, the nuclear  region of NGC 1068 presents   
a lot of emission structure  as seen in the colour map.  The average  colour around the centre is $J$ -- 2.42-$\mu$m $\sim 3$, and  gets progressively bluer with increasing distance; at $\sim$ 150 pc radius, the average  colour $J$ -- 2.42-$\mu$m  is $\sim$ 0.7.  Taking this value as  intrinsic to the central region of the galaxy, the  inferred extinction in the surrounding of the nucleus is   $A_V \sim$10--12 mag (Table 2).  For comparison, the extinction derived from the depth of the silicate feature in the VLTI / MIDI correlated spectrum is $A_V \sim 7$  mag for the more external, 3 x 4 pc cooler component, in rather good agreement with the extinction in the nuclear surrounding. Conversely, the extinction derived  for the hotter parsec-scale inner component is $A_V\sim$ 30 mag (Raban et al. 2009). \\


{\it NGC 1097}\\

NGC 1097 is   one of the nearest LINER / Seyfert type 1 galaxy in the Southern Hemisphere, at a comparable  distance as NGC 1068. The adopted scale is  1 arcsec $\sim$ 70 pc (distance = 14.5 Mpc, Tully 1988). The   nucleus is visible at all wavelengths  from UV to radio. In the IR, it is unresolved down to the highest resolution achieved in this object with VLT / NACO: FWHM $\sim$ 0.15 arcsecs  in $L$-band ($<$ 11 pc).

The SED is shown in Fig. 1. At radio waves, the following measurements are included in the SED: sub-arcsec   resolution  data from  VLA A-array  at 8.4 GHz (Thean et al. 2000) and at 8.4 GHz, the latter with a beam resolution of $0.66 x 0.25 arcsec^2$ (Orienti \& Prieto, 2009),   VLA B-array data at  1.4 GHz with beam resolution of $2.5\times1.5 arcsec^2$ (Hummel et al. 1987), and VLA B-array    archived data   at  15 GHz re-analysed in Orienti \& Prieto (2009),  and for which a final beam resolution of  1.15 $\times$ 0.45 $arcsec^2$ is obtained. Despite the  relatively larger beam of  VLA B-array data, we believe the associated   nuclear fluxes to be  fully   compatible with those  derived with the  finer scale VLA A-array data. This is based on  the analysis of  equivalent  VLA A- and B-array data at 8.4 GHz which yielded same nuclear fluxes, indicating that the nucleus is unresolved.

The mid-IR is covered with diffraction-limited resolution data from the VLT / VISIR images at 11.88 and 18.72 $\mu$m (Reunanen et al. 2009). The near-IR,  1 to 4 $\mu$m,  is covered with  VLT/NACO  adaptive 
optics images at $J$-, $H$-, $K$- and $L$- bands,  which have    spatial resolutions FWHM $< \sim$ 0.2 arcsec (Prieto et al. 2004, and this work). 

Optical and UV data were extracted from archival  \textit{HST} / WFPC2 F220W and F555W images and  
\textit{HST}/ACS F814W  image. Further UV data available from \textit{IUE} and \textit{GALEX} were not considered as their spatial  resolution beams,  20 and 6 arcsecs FWHM respectively include emission from the prominent NGC 1097's  starforming ring located at 5 arcsec radius from the centre. 

In the X-rays, the absorption-corrected  fluxes derived by Therashima et al. (2002) in the  \textit{ASCA} 0.1--4 keV  and 2--10 keV  windows are used. Further  observations at higher  energies were not found in the literature.

 For comparison purposes, the SED includes large aperture data, for the  mid-IR to the millimetric  wavelengths,   from \textit{IRAS},
\textit{Spitzer} and SCUBA, all taken from Dale et al. (2007).

NGC 1097 has shown variability in the optical, exhibiting broad emission lines. To our knowledge, there is no report on  variability  at any other spectral range.

Using the VLT / NACO colour maps $J$--$H$ and $H$--$K$,  Prieto, Maciejewski \& Reunanen (2005) estimate a moderate extinction  of   $A_V\sim$ 1 mag towards the centre.  To estimate this extinction, the colours in the  surrounding of the nucleus were  compared with those   at further locations,  inward   the circumnuclear star forming ring and not contaminated by the nuclear  dust filaments.  These filaments are readily seen in the NACO  $J$--$K$ colour map shown in Fig. 2.  \\

{\it NGC 5506} \\

This is a  Seyfert type 1.9 nucleus   in an edge-on disk galaxy, and  largely covered by dust lanes.
 As it is found in our VLT / NACO adaptive optics images, the  nucleus  dominates 
the galaxy light at IR wavelengths from 1 $\mu$m onward, but there is no equivalent counterpart in  HST optical images (Malkan et al. 1998). In the 1  to 20 $\mu$m range, the nucleus  is  unresolved  down to the best spatial resolution achieved with the  NACO observations, that is    the  $K$-band which sets an upper
limit for the size of the source of FWHM $\sim$ 0.10 arcsecs ($<$13 pc).
The adopted scale is 1 arcsec $\sim$ 126 pc (redshift taken from NED).

The SED is shown in Fig. 1. VLBA   maps  show the  nuclear region  resolved in three blobs. In the SED, the emission  from the brightest and smallest of the three blobs, also with the flattest spectral index   ($\alpha$ = + 0.06), called BO component in Middelberg et al. (2004), is  taken into account. Reported peak values from VLBA and VLBI  at 8.3, 5 and 1.7 GHz, with some of them taken at multiple epoch, are all   included in the SED. 
In addition, PTI (Parkes Tidbinbilla Interferometer) data at 2.3 GHz from Sadler et al. (1995) from a 0.1
arcsec beam is also included. 

In the mid-IR,  the extracted nuclear fluxes from VLT / VISIR diffraction-limited data at 11.8 and 18.7
$\mu$m from Reunanen et al. (2009) are included.   Additional fluxes at    6 and 9.6 $\mu$m  were directly measured on    the  6 --13 $\mu$m  spectrum published in Siebenmorgen et al. (2004, their Fig. 15), which combines ESO/TIMMI2 and  ISOPHOT   data.  Although the ESO/ TIMMI2 data correspond to a
1.2 arcsec slit-width, it perfectly joins the large aperture ISOPHOT spectrum, thus the measured fluxes should be rather genuine of the pure nuclear emission. The near-IR
 1 -- 4 $\mu$m data     are extracted  from the VLT / NACO adaptive optics images. As the images are dominated by the central source with bare detection  of the host galaxy, the nuclear fluxes were integrated within aperture sizes of 0.5 arcsec in diameter. 

Below 1 $\mu$m, the nucleus is undetected, an upper limit in the $R$-band
derived from \textit{HST} / WFPC2 archive images is set as a reference in the SED. In the X-rays,  \textit{INTEGRAL} fluxes in the 2 -- 100 keV range from  Beckmann et al. (2006) are included. The soft X-rays, 0.2 -- 4 keV,  are covered with \textit{Einstein} data (Fabbiano et al. 1992).
  
For comparison purposes, large aperture data from  the mid- to far- IR, collected with  \textit{IRAS}  (Sanders et al. 2003) are included in the SED.

There is no apparent  nuclear variability in the IR over a time-scale of   years. This follows from  the  existing agreement  between ISOPHOT, \textit{IRAS}, ground based spectra taken in 2002  (Siebenmorgen et al. 2004), and VLT/ VISIR data taken in 2006, all furthermore  having very different spatial resolution. The nucleus is however highly variable in the X-rays by  factors of 
$\sim$ 2 in scales of a few minutes (Dewangan \& Griffiths 2005).

Fig. 2 shows a  VLT / NACO $J$--$K$ colour image of the central 2.5 kpc region.  The nucleus and diffraction rings are readily seen, these  are further  surrounded  by a diffuse halo sharply declining in intensity.  
Taken as a reference  $J$--$K$ $\lesssim1.8 $ as the average  colour  in the outermost regions in this halo,   $\sim$ 150 pc  radius, and  $J$--$K$  $\sim 2.8 $ that  in the surrounding of the nucleus,  the comparison of both yields a relative  extinction towards the centre of  $A_V \sim>$5 mag. 
Due to the faintness of the galaxy, the true  extinction around the nucleus  might be  much higher than that. Most probable the colour of the halo is  largely affected by the nucleus PSF wings. For comparison, the extinction derived from the  depth of  the silicate feature at 9.6 $\mu$m in Siebenmorgen's et al. 1.2 arcsec slit-width spectrum is $A_V\sim$ 15 mag (Table 2).  \\



{\it NGC 7582} \\

This is  a Seyfert  type 2 nucleus surrounded by a ring of star forming regions. The  East side of the galaxy is largely obscured by dust lanes. These fully obscure   the  nucleus and many of the  starforming regions  at optical wavelengths.  Most of them and a very prominent  nucleus   are  revealed in  seeing-limited  VLT / ISAAC  near-IR images (Prieto et al. 2002). The achieved spatial resolution in the  current adaptive optics images allow us for very accurate astrometry on the position of the nucleus, by taking as a reference the star forming regions identified in  both the IR and HST optical images. In this way, a weak  optical counterpart source at the IR nucleus location  is found, along with a  rich network of new star forming regions, some  as close as 0.5 arcsec (50 pc) from the centre, the furthest being seen up to about  320 pc radius. The ages and masses of these regions are  analysed in Fernandez-Ontiveros et al. (in preparation). 
 The nucleus is unresolved down to the best resolution achieved in these observations, which yield a FWHM $\sim $ 0.1 arcsec ($<$ 11 pc) at 2 $\mu$m. The adopted scale  is 1 arcsec $\sim$ 105 pc (taking the redshift from NED). 

The SED is shown in Fig. 1.  The published radio maps at  8.4 and 5 GHz obtained with the  VLA-A array   show a  diffuse nuclear region  (e.g. Thean et al. 2000). 
In an attempt to improve the spatial resolution, an unpublished set of VLA-A array data at those frequencies  was retrieved from the  VLA archive and analysed by filtering out the antennas that provide the lowest resolution. In this way,  a nuclear point-like source, and some of the    surrounding star forming knots  could be disentangled from the diffuse background emission. The final beam  resolution corresponds to a   FWHM = 0.65 $\times$ 0.15 arcsec$^2$ at 8.4 GHz, and  0.96 $\times$ 0.24 arcsec$^2$ at 5 GHz (Orienti \& Prieto 2009). The radio data used in the SED corresponds to the nuclear fluxes extracted from these new radio maps. There is no further high resolution radio data   available for this galaxy. \\

Mid-IR nuclear fluxes are taken  from the analysis done on diffraction-limited VLT/VISIR images at 11.9 and 18.7
$\mu$m  by Reunanen et al. (2009).
Additional measurements in the 8 -- 12 $\mu$m  range are extracted  from
an ESO / TIMMI2 nuclear spectrum  taken with an 1.2 arcsec  slit width (Siebenmorgen et
al. 2004). Although within this slit-width the contribution from the nearest circumnuclear star forming regions is certainly included, the derived fluxes follow the trend defined by the higher spatial resolution VISIR and NACO data, which indicates the  relevance  of the AGN light  within at least 50 pc radius (0.6 arcsec) from the centre.  In the near-IR, 1 -- 4 $\mu$m, the nuclear fluxes
are extracted   from a \textit{HST} / NICMOS $H$-band image, and the VLT / NACO- narrow-band images at
2.06 $\mu$m and 4.05 $\mu$m, and  $L$-band image, using aperture diameters of 0.3 arcsecs. This aperture is  about twice the average FWHM resolution obtained in the NACO images.  In the optical, NGC 7582's nucleus becomes very absorbed.
We find a weak optical counterpart to the K-band  nucleus in  the \textit{HST}/ WFPC2 F606W image  (Malkan, Gorjiam \& Tam 1998).  The estimated flux was derived by integrating the emission in an aperture size of 0.3 arcsecs in diameter centered at the location of the IR nucleus. \\

A bright nuclear source  becomes visible again at the  higher energies. The nuclear fluxes included in the SED are extracted    from   \textit{SAX} data  in the 10 -- 100 keV band (Turner et al. 2000),  and   XMM data in the 2 -- 12 keV  (Dewangan \& Griffiths 2005). In the latter case,      the reported absorption-corrected flux is used.  An additional soft X-ray flux is  extrated from the ASCA 0.25 - 2 keV band integrated flux, not corrected by absorption, reported by Cardamone, Moran \& Kay ( 2007). 


For comparison purposes, large aperture data in the mid- to far- IR are   from  \textit{IRAS} (Sanders et al. 2003)
are also included.

NGC 7582's nucleus has shown variability in the optical, exhibiting broad emission lines. It is  also variable  in the X-rays by factors of  to  2 in intensity on scales of moths to years (Turner et al. 2000). There is no reported  variability in the IR.  On the basis of the mid-IR data used in this work, the inferred nuclear fluxes  from   TIMMI2 and VISIR observations collected four years  are  in excellent agreement.

An  $H$--2.06  $\mu$m colour map is shown in Fig. 2. This is constructed from
\textit{HST}/NICMOS $H$-band and a VLT / NACO narrow band image at  2.06 $\mu$m. The central  kpc region  shown in the map reveals   multiple star forming knots interlaced with dust.  The average colour within a radius of 180 pc from the centre and outside the star forming knots is $H$ -- 2.6 $\mu$m $\sim$ 1.6. Further out, beyond 400 pc radius and avoiding the dust filaments, the average colour is
 $H$-- 2.02$\mu$m  $\sim$ 0.9. The relative  extinction towards the centre is estimated  $A_V \sim$ 9 mag (Table 2).
The depth of the silicate feature at 9.6 $\mu$m points to larger values: $A_V \sim$ 20 mag (Siebenmorgen et al. 2004). \\



{\it NGC 1566}\\

NGC 1566 is a Seyfert type 1 nucleus  at a distance of 20 Mpc (Sandage \& Bedke 1994). Accordingly, the adopted scale is   1 arcsec  $\sim$ 96 pc.  
The HST/ WFPC2 images (e.g. Malkan, Gorjiam \& Tam 1998) show   dust lanes circumscribing the nuclear region.  Some of them can be followed up to the centre, where they seem to bend and  spiral about. The near-IR VLT / NACO diffraction-limited images reveal a smooth galaxy bulge  but some of the innermost dust lanes still leave their mark even at 2 $\mu$m. Both, at near- and mid- IR wavelengths, the nucleus  is unresolved down to the best resolution achieved,  FWHM  $< 0.12~ arcsec $ in $K$-band (11 pc). 

The SED  is shown in Fig. 1. The  high spatial resolution radio data  are taken form  ATCA observations at 8.4 GHz with a beam resolution FWHM = 1.29 $\times$ 0.75 arcsecs (Morganti et al. 1999), and  from  PTI  at 2.3 and 1.7 GHz with a beam resolution FWHM $<$0.1 arcsecs (Sadler et al  1995).
In the mid-IR, the nuclear fluxes are extracted from VLT / VISIR diffraction-limited images  at  11.9 and 18.7 $\mu$m (Reunanen et al. 2009). In the near-IR, those are extracted 
from VLT/ NACO  adaptive optics images  in  $J$- , $K$- and $L$-bands, within an aperture diameter of 0.4 arcsecs.
UV   fluxes, in the 1200 to  2100 A range,   were directly measured on  the re-calibrated pre-COSTAR \textit{HST}/FOS nuclear spectra published by Evans \& Koratkar (2004).  These fluxes were measured on the best possible  line-free spectral windows. FOS spectra were collected with the circular 0.26 arcsec  aperture diameter. Optical nuclear fluxes  were extracted from \textit{HST}/WFPC2 images with the  filters F160BW, F336W, F547M, F555W and F814W. 
 The X-ray data are from \textit{BeppoSAX} in the 20 -- 100 keV range (Landi et al. 2005), and from \textit{Einstein}  in the 0.2 -- 4 keV range (Fabbiano et al. 1992).  In the latter case, an average of the two reported measurements is used.
Large aperture data in the mid- and far- IR are from \textit{IRAS} (Sanders et al. 2003), the flux at 160 $\mu$m is from \textit{Spitzer} (Dale et al. 2007).

This nucleus may be   variable by about 70\% in the X-rays (see Landi et al. 2005) but  seems  quieter in the optical and the near-IR. Variability by a factor of  at most 1.3 over a 3 year monitoring period is reported in the near-IR, with the optical following a similar pattern (Glass 2004).\\

A VLT / NACO $J$--$K$ colour map is presented in Fig. 2. On the basis of this map, the average colour   in the  surrounding of the nucleus is  $J$--$K$ $\sim $ 1.5, the colours get bluer with increasing distance and  at about 300 pc radius, the average colour in regions outside the dust filaments is   $J$--$K$ $\sim$ 0.3. The comparison of both  implies an  extinction towards the nucleus of  $A_V \sim$ 7 mag (Table 2). \\

{\it NGC 3783}\\

NGC 3783 is a  Seyfert type 1 nucleus in a SBa galaxy. The adopted scale is 1 arcsec $\sim$ 196 pc  (redshift taken from NED). The optical  \textit{HST} ACS images show a prominent nucleus surrounded  by fingers of dust (also seen in the HST/WFPC2 F606W image by Malkan et al, 1998), a bulge and a star forming ring at about 2.5 kpc radius from the centre. In the near-IR VLT /NACO images, the emission is dominated by an equally prominent nucleus within   a  smooth and symmetric bulge; the star forming ring just get outside the field of view of these images. An upper limit to the size of the nuclear region is FWHM  $<0.08$ arcsec (16 pc) at 2  $\mu$m. First results with   VLTI / MIDI interferometry with spatial resolutions $\sim 40 mas$ at 12 $\mu$m  indicate a partially resolved nuclear region (Beckert et al. 2008). Modelling of these data with a clumpy dusty disk indicates a central structure $\sim 14 pc$ in diameter at 12 $\mu$m. This result requires further confirmation  with different base-line configurations.

The SED  is shown in Fig. 1. It includes VLA data at 8.4 GHz (FWHM $\leq 0.25$ arcsec, Schmit et al. 2001), at 4.8 GHz (FWHM $\sim $ 0.4 arcsec, Ulvestead \& Wilson, 1984) and  at 1.4 GHz (FWHM $\sim 1 $ arcsec$^2$, Unger et al. 1987).
VLT / VISIR diffraction-limited data  at  11.9 and 18.7 $\mu$m  from Reunanen et al. (2009) cover the mid-IR range.  In the near-IR, 1 to 4 $\mu$m,  the nuclear fluxes are extracted from our VLT / NACO adaptive optics images  in the J-, K- and L- bands using  apertures of $\sim$ 0.4 arcsec diameter. In the optical, nuclear-aperture photometry was extracted from archived  \textit{HST} ACS images taken with the filters F547M and F550M and the WFPC2 image with the filter F814W. The UV range is  covered  with archival 
\textit{HST} /STIS spectra in the 1100--3200 A region. Continuum fluxes were measured  on best possible  line-free regions. As NGC 3783's nucleus is very strong,  we decided in this case  to supplement the SED  with additional UV measurements   from larger aperture data, specifically, from the  \textit{FUSE} spectrum published in Gabel et al. (2003) - we measured a data point  at $\sim$ 1040 A  on their estimated  continuum flux level in their Fig 1 - in  a $U$-band measurement in a 9.6 arcsec aperture from  Las Campanas 0.6 m telescope reported in  McAlary et al. (1983).  

In the high energy range, nuclear fluxes at specific energies were extracted from   observations with OSSE  in the 50 -- 150 keV range (Zdziarski, Poutanen \& Johnson 2000),  \textit{INTEGRAL}  in the 17 -- 60 keV  (Sazonov et al. 2007),    \textit{XMM}   in the 2--10 keV range - in the later case, the reported flux  from the summed spectrum  of several observations  in Yaqoob et al. (2005) was used - and Einstein, in the 0.2 -4 KeV (Fabbiano et al. 1992).

Large aperture data in the mid-to-far IR  are from \textit{IRAS} (Moshir et al. 1990). 

NGC 3783 is known  to be  variable in the optical and the near-IR   by factors of up to 2.5 in intensity on the scale of months (Glass et al. 2004); and on  scales of minutes by a factor of 1.5  in the X-rays (Netzer et al. 2003). 

A VLT / NACO  $J$--$K$ colour map  of the central 1.5 kpc region is shown in Fig. 2.  The nucleus is the most prominent feature,  this appears  surrounded by diffraction rings and atmospheric speckles. The   colour distribution is rather flat across the galaxy:  J - K $\sim < 1$. The  inferred extinction  towards the nucleus is moderate: $A_V <$ 0.5    mag.
In the VLTI / MIDI interferometric spectrum the silicate feature at 9.6 $\mu$m  is absent  (Beckert et al. 2008).\\

{\it NGC 7469}\\

This  type 1 nucleus is the most distant source  in the sample, 20 times further than Cen A. The adopted scale is  1 arcsec $\sim$ 330 pc (from the redshift taken from NED). The nucleus is surrounded by a starforming ring that extends from 150 pc to about 500 pc  radius (as  seen in  the HST /ACS UV image in Munoz-Marin et al. 2007).  The nucleus is unresolved at near-IR wavelengths down to best resolution achieved with VLT / NACO adaptive optic images. That was in the H-band, from which an upper limit to the nucleus size   FWHM $ < 0.08 $ arcsec (26 pc) is derived.
 
The SED for this source is shown in Fig. 1. This nucleus is known to have undergone an high state level in the optical - UV in the  period  1996 --  2000,  followed by  a  slower return to a  low state level reaching a minimum  at  2004.  Changes in the continuum intensity up to a  factor of 4 were measured (Scott et al. 2005). The nucleus is also variable in the X-rays by a factor of 2.5 (Shinozaki  et al. 2006). On the other hand, monitoring of the source with IRAS pointed observations in 1983 in a time period of 22 days indicates a stable source at the 5\% level (Edelson \& Malkan 1987).
Accordingly, special attention was paid in selecting   data the most contemporaneous  possible, with most of them being taken from  the year 2000 on. Thus,  the SED is based on the following data sources. 
 
 The radio regime is covered with PTI data at  1.7 and 2.3 GHz,  with beam size of $\sim 0.1 arcsecs$ (Sadler et al. 1995), MERLIN data at 5 GHz, with beam size of $< 0.05 arcsecs$ (Alberdi et al. 2007) and VLA archived data at 8.4 and 14 GHz, reanalysed in Orienti \& Prieto (2009), and for which  resolution  beams of $< 0.3~ arcsec$  and $< 0.14 ~arcsec$ respectively are obtained.

In the IR, nuclear fluxes were extracted from VLT / VISIR diffraction-limited images  at  11.9 and 18.7 $\mu$m collected in 2006 (Reunanen et al. 2009), 
 VLT/ NACO  adaptive optics images in  $J$-, $H-$, $K-$- and $L$-bands and  the narrow-band continuum filter at 4.05  $\mu$m, all collected from 2002 on (Table 12),  and the HST / NICMOS narrow-band continuum image at 1.87  $\mu$m, collected in 2007.
 
 In the optical and UV, the nuclear fluxes were extracted from the HST /ACS  F330W (collected in  2002),  F550M and F814W (both in 2006), and WFPC2  F218W (1999) and F547M (2000) images.  The aperture size used in all cases,  VLT-IR and HST-optical-UV,  was 0.6 arcsec in diameter.  This selection was a compromise  between getting most of the  light in the nuclear PSF wing and  avoiding  the star forming ring.  

In the X-rays,   nuclear fluxes at specific energies were extracted from   observations with INTEGRAL in the 17 - 60 keV band  (Sazonov et al. 2007), XMM  in the 2-10 keV band (Shinozaki el al. 2006) and ROSAT in the 0.1 - 2.4 keV band  (Perez-Olea \& Colina 1996). In the case of XMM and ROSAT, the fluxes included in the SED  were derived from the luminosities provided by the  authors, and we thus assume they are intrinsic to the source, although this is not explicitly mentioned to be the case.
    The ROSAT flux is nevertheless not genuine from the nucleus as it includes  the contribution of the complete star forming ring.\\

The SED is complemented with large aperture data in the IR, taken from IRAS (Sanders et al. 2003) and Spitzer (Weedman et al. 2005), and millimetre from SCUBA (Dunne et al. 2000) and  the Caltech Submillimeter
Observatory (Yang \& Phillips 2007).

A VLT / NACO  $J$--$H$ colour map  of the central 4 kpc region is shown in Fig. 2. This was selected instead of the usual $J$--$K$ because of the better spatial resolution reached in the H-band. The map shows the nucleus and a ring of diffuse emission which just encloses the star forming regions, with radius of 500 pc. Further out from the ring, the signal in the individual near IR images drops dramatically, and 
 a reliable estimate of the intrinsic galaxy colours is not possible. Thus, an estimate of  the relative extinction around the 
nucleus is not provided in this case. 
As in NGC 3783, the available VLTI /MIDI interferometric  spectra do not show evidence for a silicate feature.

\section{Comparing with a Quasar: the SED of 3C 273 } 

 3C 273 is one of the most luminous quasars in the sky reaching  a  luminosity of several 
$10^{46}$ erg sec$^{-1}$ at almost any energy band.
In gamma-rays,  its luminosity  reaches $10^{47}$ erg sec$^{-1}$,  which  is suspected to be due to strong beaming at these energies. 

3C 273 is a radio loud source while all  the other  AGN discussed in this work are radio quiet, with the possible exception of Cen A.  It is also the most distant object in the sample: the adopted scale is 1 arcsec $\sim$ 3.2 kpc  (redshift taken from NED). The high power of 3C 273   makes  its SED insensitive to the  spatial resolution data used  at almost any band, except in the low frequency regime where some of the
jet components are somewhat comparable to the core strength. 
That  together with  its excellent  wavelength  coverage lead us  to use this SED  as  a reference for a no obscured AGN. 3C 273  is variable mainly at  high energies by factors of  3 to 4. But from optical to radio,  the reported variability  is  20 to 40\% (Lichti et al. 1995; Turler et al. 1999),  which  might affect the SED  in detail  but has minimal impact on   the  overall  shape.
 The SED of 3C 273 is shown in  Fig. 1. The data  from  1 $\mu$m  up to  the gamma-rays are  from Turler et al. (1999). These data are indeed an  average of multiple observations at distinct epochs. Further in  wavelength, the following set of data are used: for consistency with the rest of the work presented here, the data beyond  1 $\mu$m up to 5 $\mu$m  are taken from our VLT / NACO adaptive optics images (collected in May 2003); the  difference with the larger aperture data used in  Turler et  al.  is less than  10\%. The  6 to  200 $\mu$m range is covered with  \textit{ISO} (collected in 1996, Haas et al, 2003). In the  millimetre  and    radio waves we used higher resolution data than that in Turler et al. to better isolate the core from the jet components. Those are: VLBI observations at 3 mm   from Lonsdale, Shepher \& Robert (1998) and 147, 86, 15 and 5 GHz  from  Greve et al. (2002), Lobanov et al. (2000), Kellerman et al. (1998) and  Shen et al.  (1998) respectively. Additional VLBA measurements   at 42 and 22 GHz - peak values - are taken from Marscher et al. ( 2002).

A VLT /  NACO $J$-$K$ colour map is shown in Fig. 2. This just  shows  the central point-like source and  various diffraction rings. 

The SED of 3C 273 differs from all those shown in this work in  the radio domain mainly, presenting a flatter spectrum as expected from    a radio loud source.    In the optical to UV,  the difference is also important with type 2 nuclei because of the dust absorption in the latter but less so with type 1s.
For comparative purposes, an artificial extinction of $A_V = 15$ mag was applied to the SED of  3C 273  and the result is shown on top of its  SED  in Fig. 1.   With extinction values in this range, which is on average what  we measure from the near-IR colour maps in the galaxy sample,  the UV to optical region in 3C 273 becomes   fully absorbed,  presenting a closer resemblance with   that shown by the  Seyfert type 2 nuclei. 

The comparison of 3C 273 with  the type 1 nuclei is  more illustrative. The average of the  three genuine type 1 SEDs  compiled in  this work is shown together with that of 3C 273 in  Fig. 3. The  SEDs are     normalised to the mean value  of their respective power distributions, in this way  they appear  at comparable scale about the optical region. All the SEDs show in the optical to UV the characteristic blue bump feature of type 1 sources and the usual inflexion point at about 1 $\mu$m. However, whereas the blue bump is  stronger in 3C 273, indeed the dominant feature in the UV to IR region,  that is weaker and softer in energy in the Seyfert 1's.
Conversely,   in the IR  the  type 1 objects show  a broad    bump feature  in the 10  - 20 $\mu$m range whereas 
3C 273 shows a    flatter distribution over the same spectral region, the general trend being of a shallow  decrease in power  with decreasing frequency.  \\

  A weaker UV bump is an indication of more dust in the line of sight to these type 1 nuclei, but  the lack of the IR bump in 3C 273,  which is a key diagnostic of   the existence of a central obscuring dust structure  in any AGN,  points to no much  dust  in this nucleus.  This absence may still be  due to a strong non-thermal contribution  - this is a flat-spectrum  radio source -  which would smear out any IR bump. However,  this will make the dust contribution to the IR even lower. Some  hot dust  may still exists in 3C 273, as traced by the detection   of silicate features in emission at  10 and 20 $\mu$m (Hao et al. 2005). Thus, the 
overall evidence   points to a poor dusty  environment in this object as compared with those of lower luminosity AGN. 
There are also first results with VLTI / MIDI and Keck interferometry on 3C 273 which indicate a slightly resolved nuclear structure at 10 and 2 $\mu$m respectively (Tristram et al. 2009 and Pott 2009). However it is more plausible that this structure is caused by  the jet of 3C 273    rather than by dust. 

At the high  energies, the comparison of the Seyfert's SED shape with that of 3C 273 is limited by the poorer spectral coverage of the former.  Still,  in the overlapping region, from the soft X-rays till $\sim $200 keV, the SED of all the Seyfert's  present     a   gentle rise in power with increasing frequency. This trend may point out to   the existence of a further emission bump   at much  higher energies  but  that may escape detection with present facilities.  A broad emission bump peaking at the MeV region is detected  in both  3C 273 and Cen A, both with jets detected in the X-rays.

\begin{figure}[p] 
   \centering
 \plotone{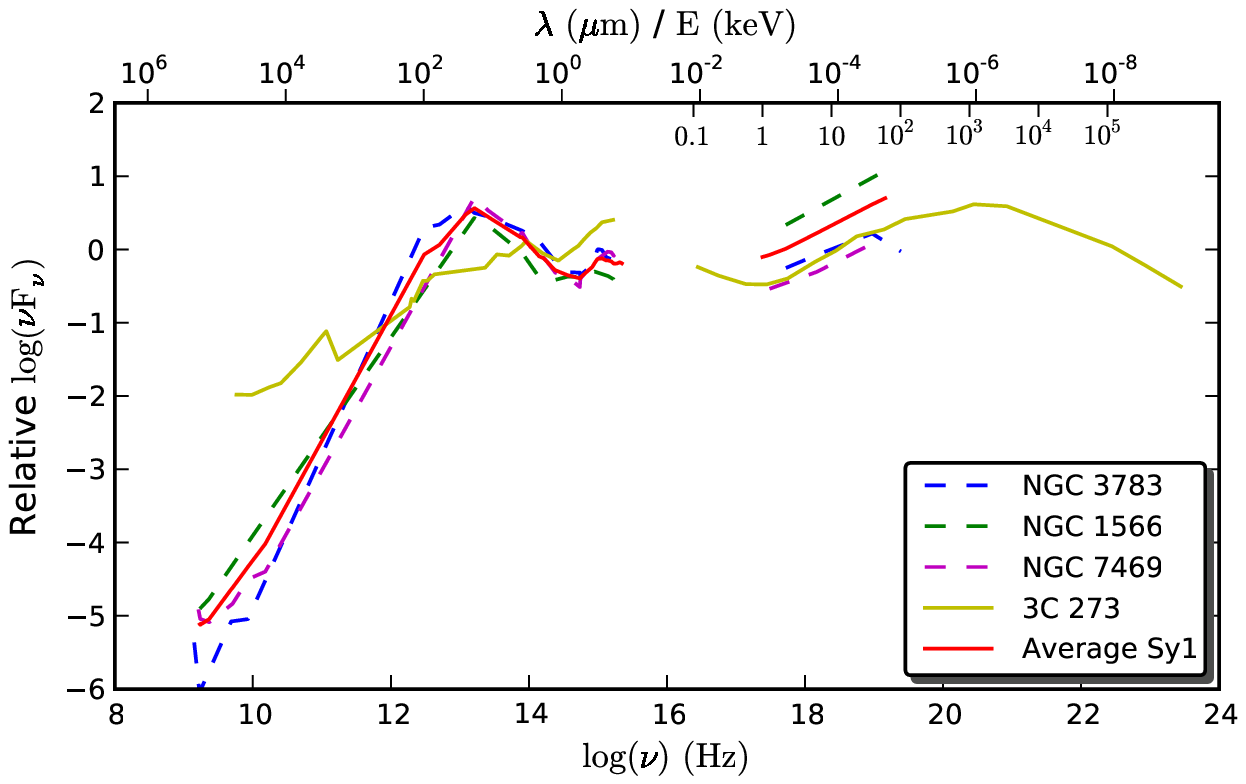}
   \caption{Average SED - in red - of the type 1 nuclei in this work: NGC 1566, NGC 3783 and NGC 7469.
 3C 273 - in yellow - is shown for comparison.  Prior averaging,  each SED was set to its  rest frame system  and then  normalised  to the mean value of its $\nu F_\nu$ distribution.  In the X-rays, the   average is determined in the  common to the three objects energy-window, 1 to $\sim$ 60 keV. }
   \label{Fig. 3}
\end{figure}

\section{Results: the new  SED  of nearby AGN }
\subsection{The SED shape}

The high spatial resolution SEDs (Fig. 1) are all characterised by two main features: a emission bump in the IR and an increasing trend in power at the high energies. 

The available high spatial resolution data  spans  the UV,   optical, and IR  up to about 20 $\mu$m. Shortward of $\sim 1 \mu$m, all the type 2 nuclei are undetected, or barely detected (e.g NGC 7582). Longward of 20 $\mu$m, there is a wide data gap up to the radio frequencies due to the lack of data of comparable spatial resolution.  Subject to these limitations, all type 2 nuclei are characterised   by  a sharp decay in power  from 2 $\mu$m onwards  to the optical wavelengths. Conversely, 
type 1 objects present also a  decay shortward of  2 $\mu$m but this   recovers  at about 1 $\mu$m  to give rise to the characteristic  blue bump feature seen in Quasars and type 1 sources in general  (e.g.  Elvis et al. 1994).
 The inflexion point at  about 
1$\mu$m is also a well known feature in AGN, generally ascribed to a signature of dust emission at its limiting sublimation temperature (Sanders et al. 1989). The only LINER in the sample, NGC 1097,   appears as an intermediate case between both type 1 and 2: it is detected up to the  UV wavelengths but  there is no blue bump, its overall SED being more  reminiscent of  a type 2 nucleus.

Longward of 2 $\mu$m  all the SEDs tend to flat and there is some hint of a turnover towards lower power at about 20 $\mu$m in some objects. The large  data gap     in the far-IR to the millimetric  wavelengths leaves us with  the ambiguity on   the   exact shape  of the IR bump and its  width.  Because of  the small physical region sampled in the near- to mid- IR,  on the scale of   tens of parsec, an important contribution from cold    dust at these radii    that would     produce a secondary IR-millimetre bump    is not anticipated. Instead, a smooth  decrease in $\nu F_\nu$ towards the radio  frequencies is expected.  This suggestion  follows from the   SED turnover beyond 20 $\mu$m shown by   the  objects for which  large aperture data in the far-IR or the  millimetre wavelengths could be included in their SEDs, namely NGC 3783, NGC 5506 and Cen A.
 
The complete galaxy sample is  detected in the 0.2 --100 keV region - with the exception of  NGC 1097 for which no reported observations beyond 10 keV  were found. All show a  general  increase in  power with frequency.  At  higher energies, Cen A is the only source detected up to the MeV region. Indeed, among   low-power AGN, Cen A is so far the only source      detected at gamma-rays (Schonfelder et al. 2000). 3C 273 is of course detected at these energies and as Cen A, both exhibit   a rather broad bump peaking in the  MeV region. 

Fig. 4 shows the  average of the type 2 SEDs - excluding the case of Cen A because of its non-thermal nature,  see sect. 7 - compared with that of the type 1's (same  average SED   shown   in Fig. 3). The same procedure to produce the average type 1's is used for type 2's: each  type 2 SED is normalised to the mean of its power distribution,  the resulting SEDs    being then averaged. The resulting average  template for each type are  plotted one on top of the other in the figure.  It can be seen 
that the  most relevant feature in both  SED types is the IR bump. This can be reconciled with emission from dust with an equivalent grey-body temperature of $\sim$ 300 K in average (section 5.2 ).  The location and shape of the bump  longward of 
2~$\mu$m is similar for both types; it is shortward of this wavelength  where the difference arises: type 1s present a shallower 2 $\mu$m-to-optical spectrum which is further  ensued by the blue bump emission.   This dramatic difference suggests   a clearer sight line to hot dust in  type 1s but and obscured one in type 2s. This is fully consistent with the torus model: the shallower spectrum reflects the contribution of much  hotter dust from the inner region of the torus which we are able to see directly in type 1s; in the type 2s this innermost  region  is  still fully   absorbed by enshrouding colder dust.

The second important feature in both SED templates is the high energy spectrum. Within the  common sampled energy band -- 0.1 to $\sim$ 100 keV -- this  region appears   rather similar in both AGN types,  the general trend being  that of a  gentle increase  in power with increasing frequency.

\begin{figure}[p] 
   \centering
 \plotone{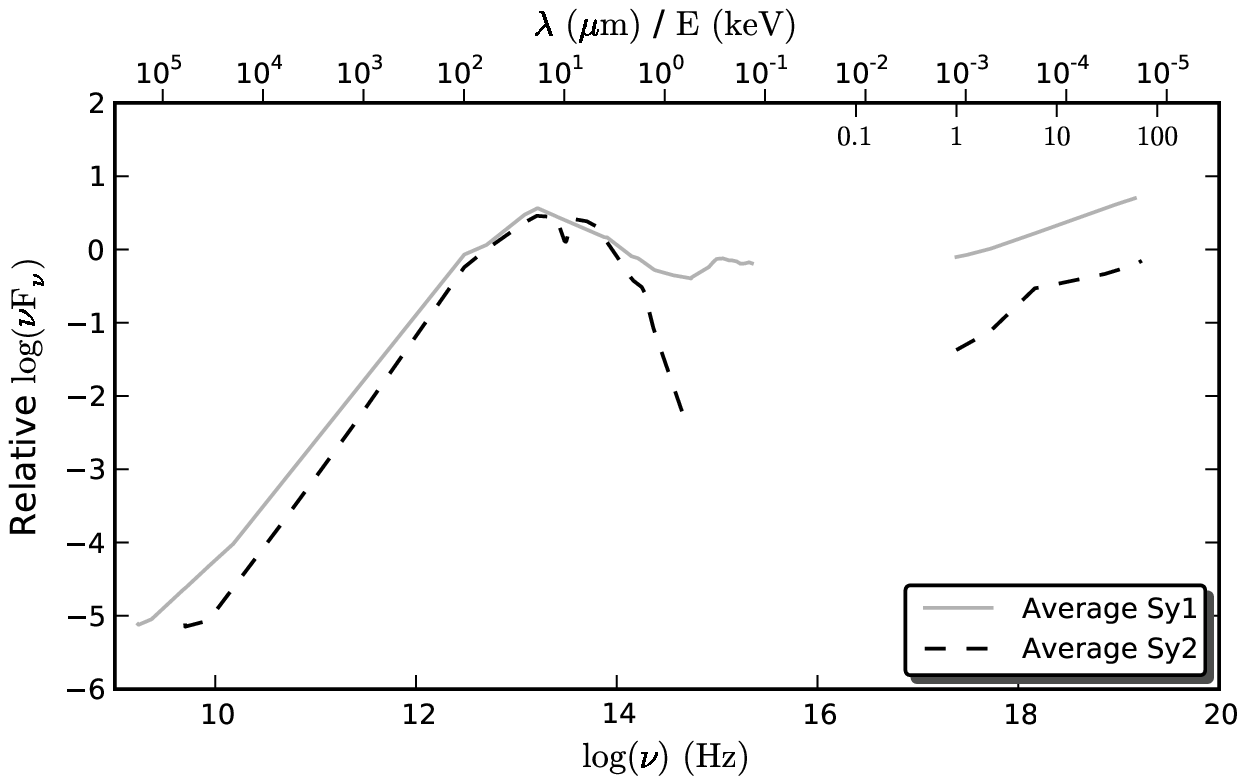}
   \caption{Average SED template of type 1s,  in  thick line - includes NGC 3783, NGC 1566 and NGC 7469 - and  of  type 2s, dash line - Circinus, NGC 1068, NGC 5506 and NGC 7582. Before averaging, each SED
  was set to its rest frame  system and  then normalised  to the mean value of its $\nu F_\nu$ distribution. The average for type 2s in  the X-rays   is determined in the common  energy window,  1 to  70 keV, that of type 1s in the energy window 1 to 60 keV.  }
   \label{Fig. 4}
\end{figure}

\subsection{Core luminosities}

Having compiled a more genuine   SED for these AGN, a tighter  estimate of
their true  energy output  can be derived by integrating the SED. 
This is done over the two main features in the SED: the IR bump and  the  high-energy  spectrum (Fig. 1). For the type 1 sources, a further integration was done over the blue bump. The estimated energies associated with each of these regions  are listed in Table 1.  
The procedure used is as follows.

 The integration over the IR bump extends from the inflection point at about 1  $\mu$m in type 1s,  from the optical upper limit in type 2s, up to the  radio frequencies. Direct numerical integration on the $F_v$ vs $\nu$ plane was done. For  those  sources whose nuclear flux  is independent on the aperture size, namely, NGC 5506, NGC 3783 and 3C 273, the integration  includes the large aperture IR data as well. For Cen A, the millimetre data were also included.  For all other sources, a linear interpolation between 20 $\mu$m and  the first radio frequency point   was applied.  Effectively, the proposed integration   is  equivalent to     integrating over the  1  -- 20 $\mu$m range only as this region dominates the total energy output by orders of magnitude in all the objects.
As a control, a  modified black body (BB) spectrum, $B_{\nu}\times\nu^{1.6}$, was  fitted to this spectral range.  The derived effective temperature converges  into the 200 - 400 K range in most cases, and the  inferred IR luminosities   are  of the same order of magnitude as those derived by the integration procedure over the SED. Focussing on  the objects for which   their nuclear fluxes are independent on the aperture size -- NGC 3783 and NGC 5506 --  a  further test was done by comparing their  integrated luminosity  in  the  1 -- 20 $\mu$m range with that in  the 1 -- 100 $\mu$m range, the latter including all  the available data.    The 1 -- 20 $\mu$m luminosity is   found about a factor of 1.5 smaller. Thus,  the IR luminosity is probably underestimated by at least this factor in all other sources.\\

At  high energies,  due to   energy-band overlapping  between   different satellites,  a direct  integration over   the  SED was avoided.  Instead,   we used X-ray luminosities reported in the literature, selecting those  derived  from the hardest  possible energy band, usually  in the 20 -- 100 keV range.
  X-ray luminosities above 20 keV are less subjected to absorption and   thus  expected to be a fair indication  of the nuclear budget (with the possible exception of  NGC 1068 due to the large X-ray column density, $N(H) > 10^{25} cm^{-2}$,   inferred in this case, Matt et al. 1997). \\


For the   Seyfert type 1 nuclei, the integration over the  blue bump  spans the 0.1 to  1 $\mu$m region.
The same integration procedures described above were applied to 3C 273 as well.
 
On the basis of these energy budgets, an estimate of the bolometric luminosity  in these AGNs is  taken  as  the sum of the IR plus  X-ray - above 20 keV - luminosities (Table 1). In doing so, it is implicitly assumed that the IR emission  is a genuine measurement of the AGN energy output  and accounts for most of the  optical to UV to X-ray luminosity 
 generated in the accretion disk.
 

%

\section{The extinction gradient towards the centre}
Table 2 gives a comparison of  nuclear extinction values inferred from different methods. 
A first order estimate  is derived from the near-IR colour maps presented in this work. With these maps,  the colours  in the surrounding of the nucleus are compared  with those  at further galactocentric radii, usually at several hundred parsecs. Colour excesses   are found to progressively increase towards the central region, an indication of an  increasing  dust  density  towards the nucleus. In some cases however the distribution of colours is rather flat, e.g. in NGC 1097,  NGC 3783.  These central   extinctions, in some cases inferred at     distances of  30 - 50 pc  from the centre, are  systematically  lower, by  factors of 2 - 3, than those inferred from the silicate feature at 9.6 $\mu$m  (Table 2 ).
Considering the very high spatial resolution of some of the silicate feature measurements,  the  difference is  an indication  that  the distribution of absorbers at the nuclear region  is not smooth but has a high peak concentration at the very centre.

A further  comparison  with the optical extinction inferred from the X-rays column density, assuming the standard Galactic dust-to-gas ratio and extinction curve  (Bohlin et al. 1978), is also given in Table 2. It is known that the extinctions derived this way are always very large. For the objects in this work, they are several  factors,  or even orders of magnitude in the Compton thick cases, higher than those inferred from the silicate feature.  Only in  NGC 5506,  the values derived  from both methods agree. Such  high discrepancies  have to be due to the inapplicability of  the Galactic dust-to-gas ratio conversion in AGN environments  (see Gaskell et al. 2004).

\section{Discussion} 

{\it The IR SED: large aperture vs high spatial resolution} \\

We have compiled spectral energy distributions at subarcsec scales for a sample of nearby well known AGN. These SEDs reveal major differences in the IR region when compared with those based on IR satellite data. First, 
  the trend defined by the large aperture IR data (crosses in  fig. \ref{Fig. 1}) is different from that defined by 
the high spatial resolution data  (filled points in fig. \ref{Fig. 1}). Second,  the true AGN  fluxes can be  up to an order of magnitude lower that   those inferred from 
large aperture data, hence  the bolometric luminosities based on IR satellites data can be overestimated by orders of magnitude.

The number of objects studied at these high spatial resolutions  is small. They are however among the nearest AGN. Subjected to this limitation, if we take the new SEDs  as a reference for the Seyfert  class,  the above results have two further implications: 
\noindent

1) the AGN contribution to  the  mid-to-far IR  emission measured by e.g.  \textit{IRAS}, \textit{ISO}, \textit{Spitzer} is minor, the  bulk of the  emission measured by these satellites comes from   the host galaxy.   This result fully confirms previous work by Ward et al. (1987),  who pointed out  the relevance  of the host galaxy light  in  the IRAS fluxes  in  already type 1 AGNs. On this basis it is understandable  the radio -- far-IR correlation   followed by  AGN  and normal star forming galaxies  alike (Sopp \& Alexander 1991; Roy et al. 1998; Polleta et al. 2007). A common  trend indicates that   the far-IR emission  is unrelated to the AGN. The shape of  the high spatial resolution SEDs of the  sample AGN   shows that the large aperture mid-IR is also unrelated in most cases. 

\noindent

2) the  selection or discrimination of AGN populations  on the basis of  mid  to far IR colours may not be applicable on a general basis. For example,  Grijp's et al. (1987) criteria based on  the   \textit{IRAS}
  density flux ratio at 60 and 25 $\mu$m,  $f_{60} / f_{25} > 0.2$,    to  find predominantly   AGN. Recently, Sanders et al (2007)  proposes the use  the colour of the   Spitzer  3.6 to 24  $\mu$m spectrum to help separating type 1- from type 2-   AGN.
Considering the   shape and luminosity of the high spatial resolution   SEDs, it  is somewhat surprising that large  aperture mid-to-far IR colours may keep track of   the existence of a central AGN in most galaxies.  \\

However, the above  criteria should apply to cases where the AGN    dominates the IR galaxy light in  a similar way as   in  quasars.  Two of  the high spatial resolution SEDs studied  reflect this situation: NGC 5506, a type 2- , and NGC 3783, a type 1-  nucleus. In both,    large aperture  data  remain genuine of the nuclear emission, as it is also the case of the quasar 3C 273 used here for comparative purposes. The  SEDs of these objects show a smooth  connection between  the large aperture far- to mid- IR data  and the high spatial resolution  mid- to near- IR data (Fig. 1).  In NGC 5506,   the   nucleus's high contrast in the IR  is due to the low surface brightness and  edge-on morphology of the  host galaxy. 
However, this is not the case for NGC 3783 which resides in a bright   face-on spiral.  3C 273 resides in an elliptical galaxy but as an extreme powerful quasar it dominates the light of its host at any wavelength.  The relevance of the AGN in these objects is  more obvious    in the $J$--$K$ colour maps shown in  Fig. 2. These are dominated by the central source, and   almost no trace of the host galaxy  at further radii from the centre are detected indicating a flat and smooth galaxy light profile. This  morphology is rather different from   what is seen in the other AGN in the sample whose  colour maps  highlight the central dust distribution. \\

NGC 5506 and NGC 3783 happen to be among the  most powerful nuclei in the sample, besides  3C 273, with IR luminosities above $ 10^{44}$ erg s$^{-1}$.  The nearest in power is NGC 1068 with 
LIR $\sim 8.6\times 10^{43}$ erg s$^{-1}$ (Table 1), a mere factor of 2 lower. 
  Thus, there is a tentative indication   that  AGN  luminosities above $10^{44}erg s^{-1}$ may  easily be traceable   in the IR regardless of the  aperture flux  used. One would expect quasars to  naturally comply with   this criterion but the generalisation is not that obvious. A third AGN in the sample, NGC 7469, has   an IR  core luminosity  of $2\times 10^{44}erg s^{-1}$, but its host galaxy dominates the total IR budget by  a factor of seven  (Table 1).\\

{\it Total energy budget: IR vs X-ray  luminosities} \\

The   derived nuclear luminosities  in the IR should  provide a tighter  quantification  of the dust-reprocessed optical to UV and X-rays photons produced by the AGN - the accretion luminosity and X-ray corona. On that premise, an estimate of     the total energy budget in these objects as  the sum of their IR and hard X-ray emissions is evaluated ( $ L_{IR} + L_{> 20 keV}$ in Table 1).  We compute the same number in type 1s as well, on  the assumption that  the blue-bump emission visible in these objects  is accounted for in the  IR  reprocessed emission  and thus is not summed up to the total budget (following similar reasoning as in e.g Vasudevan \& Fabian, 2007).  We account for  the  X-ray contribution  for energies above 20 keV as photons beyond this energy should provide  a genuine representation of the nucleus plus jet emission. 
 
 In comparing the IR and X-ray luminosities in Table 1, the  IR  luminosity  is found to dominate the total  budget, by more than $\sim 70\% $ in  seven out of the ten cases  studied. Indeed, in most of these cases, the X-ray emission is a few  percent of the total. The three exceptions include Cen A, in the border line with an  IR contribution  about 60\% of the total, and 3C 273 and NGC 1566, where the IR contribution  reduces to less than 30\% of the total. Cen A    and  3C 273 are  the two sources in the sample with a strong jet, also in the X-rays.
 
 As it can be explored from the table 1, there  is not dependance of the IR to X-ray ratio on the AGN luminosity but the sample is small. This ratio is furthermore vulnerable to variability.   As reported for each object in sect. 3, variability in the X-rays is common in these objects, by  a factor of 2 to 3 in average - up to a factor of 10 has been seen in Cen A - whereas   variability in the IR is at most a factor of 2, so far only known for Cen A, NGC 1068 and NGC 3783. Thus, X-ray variability which is also faster in time scales,   can modify the IR to X-ray ratios up and down by the same factors. Still, even if accounting for a positive increase in the X-ray luminosity by these factors, the IR luminosity remains the dominant energy output for most cases. Just  a factor of 2 decrease in the X-rays in  the three objects with a reduced IR core,    Cen A, 3C 273 and NGC 1566, will place their X-ray and IR luminosities  to the same level.

 
Focusing  on the   Seyfert type 1 objects, all characterised by   a   blue  bump  component in the SED,   the luminosity associated with the observable part  of this region is  $\sim 15\% $ of the IR luminosity, whereas in 3C 273, integrating over the same spectral region, that is  90\% (Table 1).  A  fraction of the blue bump  energy is unobserved as it falls in the extreme UV  to soft-Xrays data gap, still,  the fact that there is  almost a factor six difference in these relative emissions between  3C 273 and the Seyfert  type 1 nuclei is an indication that  Seyfert AGN  are seen  through much more dust.  This is in line with  conclusions reached by  Gaskell et al. (2004) who argue  on the  presence of additional reddening by dust in  radio-quiet  as compared with radio-loud AGNs. If this is the case,   the IR bump luminosity may be one of the most tight measurements of the accretion luminosity in Seyfert galaxies. \\

{\it Centaurus A: a special case}\\
 
Among all AGN  analysed in this work Centaurus A's SED singles out as a particular case.
The data points from  VLBA  over the millimetre to the high resolution IR measurements  follow a rather continuous trend.  This region can be fitted by  a simple synchrotron model with   spectral  index,  $F_{\nu}  \sim \nu^{-0.3}$, and still the   $\gamma$-ray emission 
 be explained as inverse Compton scattering of the radio synchrotron electrons
(Prieto et al. 2007, see also Chiaberge et al. 2001). Such a flat synchrotron spectrum 
 has been suggested by Beckert \& Duschl (1997 and references therein) for low luminosity AGN, among which Cen A nucleus can be considered. The available high spatial resolution  IR observations indicate that most of the emission comes from a very compact source  less than 1 pc in diameter (Meisenheimer et al. 2007). This together with the apparent synchrotron nature of its SED points  to a rather  torus-free  nucleus. Cen A is one of the lowest power sources in the sample,  with  $ L(IR)\sim10^{42}$ erg s$^{-1}$. On theoretical grounds, it is being argued  that  AGNs of this low power may be unable to support a torus structure and should thus show a bare nucleus at IR wavelengths (e.g. Hoenig \& Beckert 2007; Elitzur \& Shlosman 2006).
In other respects, Cen A's SED is similar to those of type 2 nuclei in this study, in the sense that its optical to  UV is totally obscured  but this  may be caused by the   large scale dust lanes crossing in front of its nucleus. The future availability of millimetre data of high spatial resolution will  help to confirm the nature of this SED. \\

\section{Conclusions} 

Sub-arcsec resolution data spanning the UV, optical, IR and  radio   have been used to construct  spectral energy distributions of the central, several  tens of parsec,  region of  some of the  nearest and brightest active galactic nuclei. Most of these objects are Seyfert galaxies. \\

These high spatial resolution SEDs differ largely from those derived from large aperture data, in particular in the  IR: the shape of the SED is different and the true AGN luminosity can get overestimated by orders of magnitude if based on IR satellite data. These differences appear to be critical for AGN luminosities below $10^{44} erg~s^{-1}$ in which  case large aperture data  sample  in full the host galaxy light.  Above that limit we  find cases among these nearby Seyfert galaxies where the  AGN   behaves  as the  most powerful  quasars,   dominating   the host galaxy light  regardless of the integration aperture-size used. \\

The high spatial resolution SED of these nearby AGNs  are all characterised by two major features in their power distribution: an IR bump  with maximum in the 2 -10  $\mu$m range, and an increasing trend in X-ray power  with frequency in the  1 to $\sim$ 200 keV region, i.e. up to the hardest energy that was possible to sample. These  dominant features are common  to  Seyfert type 1 and 2 objects  alike. \\

The major difference between type 1  and 2 in these SEDs arises  shortward of 2 $\mu$m. Type 2s are  characterised by a sharp fall-off shortward of this wavelength, with no optical counterpart to the IR nucleus being detected  beyond  1 to 0.8 $\mu$m.
Type 1s show   also a drop shortward of 2 $\mu$m but this is   more gentle - the spectrum is flatter -  and  recovers at about 1 $\mu$m to give rise to the characteristic   blue-bump feature seen in quasars. 
The flattening of the spectrum   shortward of 2 $\mu$m is also an expected feature  of type 1 AGNs. Interpreting  the  IR bump  as AGN reprocessed emission by the nuclear dust,  in type 1s   the nearest to the centre hotter dust can be directly seen, hence the flattening of their spectrum, whereas in type 2s this hot dust   is still fully obscured. \\

Longward of 2 $\mu$m, all the AGN types   show very similar SEDs, the bulk of the IR emission starts from this wavelength on and  the shape of the IR bump is very similar in all the AGNs. This is compatible with an equivalent black-body temperature for the bulk of the  dust in the 200 - 400 K range in average. Although the current shape of the IR bump is limited by the availability of high angular resolution data  beyond    20  $\mu$m  for most objects, due to  the small   region sampled in these SEDs, of just a  few parsecs in some galaxies, a major contribution from colder dust that will modify the IR bump is not expected. \\

It can thus be concluded that at the scales of  a few tens of parsec from the central engine, the  bulk of the IR emission in either AGN type can be reconciled with pure dust emission. It follows that further    contributions from   a non-thermal synchrotron component and/or  a thermal free-free emission linked to cooling of ionised gas  are insufficient to overcome that of  dust at these physical scales. 
The detailed modelling of  NGC 1068's SED - this being one of the most complete we have compiled -  in which   these three contributions --   synchrotron, free-free and a dust torus   components -  are  taking into account illustrates that premise, that is, the dominance of dust emission in the IR, even at the parsec-scale resolution achieved for  this object in the mid-IR with  interferometry (Hoenig et al.  2008).
Only the two more extreme objects in this analysis, Cen A, on the low luminosity rank, and  3C 273, on the highest, 
present a SED that is not   dominated by dust but by  a  synchrotron component. We tend to  believe that is due to a much reduced dust content  in these nuclei. \\

Over the nine orders of magnitude in frequency covered by these SEDs, the power stored in the IR bump is by far the most energetic  fraction of the total energy budget measured in these objects.  Evaluating this total budget as  the sum of the  IR  and hard X-ray -- above 20 keV --  luminosities,  the IR part accounts for more than 70\% of the this total  in  seven out of the ten AGN studied. In the three exceptions, the IR fraction reduces to $<\sim 30\% $ (3C 273 and NGC 1566), $< \sim 60\%$ in Cen A. Even if accounting for variability in the X-rays, by a factor 2 to 3 in average, the IR emission    remains in all cases dominant  over, or as  important as, in the last three cases, the  X-ray emission. 
If comparing with the  observed blue bump luminosity   in the type 1 nuclei, this  represents less than  15\% of the IR emission. Putting all together, the IR bump  energy from these high spatial resolutions SEDs  may represent the tightest measurement of the accretion luminosity in these Seyfert AGN. \\

The average high spatial resolution SED  of the type 2 and  of the  type 1 nuclei analysed in this work, and presented in Fig. 4, can be retrieved from http://www.iac.es/project/parsec/main/seyfert-SED-template.
\\

This work was initiated and largely completed during the stay of   K. Tristram, N. Neumayer and A. Prieto at the Max-Planck Institut fuer Astronomie in  Heidelberg.

\begin{deluxetable}{lcccccccccccc}
\tabletypesize{\scriptsize}
\tablecaption{Energy budget in  nearby AGN. Objects ordered by increasing IR core luminosity}
\label{sed}
\tablehead{\colhead{Name} & \colhead{FWHM$_{nucleus}$ } &
\colhead{L$_{IR-HSR}$} & 
\colhead{L$_{X 
> 20 keV}$}  & 
\colhead{$L_{total}$ }  &
\colhead{$\frac{L_{opt-UV}}{L_{IR-HSR}}$} &
\colhead{$\frac{L_{IR-HSR}}{L_{total}}$} &
\colhead{$\frac{L_{IRAS}}{L_{IR-HSR}}$} &
\colhead{L$_{IRAS}$}\\
 \colhead{ AGN type} & \colhead{in pc at 2 $\mu$m} &\colhead{erg/s} &\colhead{erg/s}  & 
\colhead{erg/s}  &
\colhead{\%} & {\%}& &
\colhead{erg/s} \\ }
\startdata

NGC 1097/ LI  & $< $11 & 3.3x$10^{41}$ & 5.2x$10^{40}$ & 3.8  x$10^{41}$ &	     -   & 87  & 500 & 1.5x$10^{44}$  \\
Cen A / T2        & $<$1 & 1.7x$10^{42}$ &  1.1 x$10^{42}$  & 2.8x$10^{42}$  &   -      & 61 &  20 & 3.5 x$10^{43}$  \\ 
NGC 1566/ T1  &  $< $11 & 1.8 x$10^{42}$  &  3.7x$10^{42}$  & 5.5  x$10^{42}$ & 14 & 33  & 78  & 1.4x$10^{44}$ \\
Circinus / T2    & $\sim$2 &  8x $10^{42}$   &  4.3x$10^{41}$   & 8.4 x$10^{42}$ &   -  & 82  &   7 & 5.5 x$10^{43}$ \\
NGC 7582/ T2   & $<11$ & 1.3x$10^{43}$ &  4.5 x$10^{42}$  & 1.8 x$10^{43}$  &   -  & 72 &  25 &  3.2 x$10^{44}$ \\
NGC 1068/ T2  & 1.2 x2.8 & 8.6 x$10^{43}$ & 6.5 x$10^{41}$ & 8.7 x$10^{43}$ &  -    & 99 &   9  & 7.8 x$10^{44}$\\

NGC 5506/ T2  & $<13$ & 1.2 x$10^{44}$ & 6.6 x$10^{42}$ & 1.3  x$10^{44}$   &   -   & 99 & 1.1  & 1.3 x$10^{44}$ \\
NGC 3783/ T1   & $ < $16 & 1.6 x$10^{44}$ & 2.5 x$10^{43}$  & 1.8 x$10^{44}$& 13 & 89  & 0.8  & 1.3 x$10^{44}$ \\
NGC 7469/ T1  & $ < $ 26 & 2.2 x$10^{44}$  & 3 x$10^{43}$  & 2.5  x$10^{44}$ & 10 &  88   &  7   & 1.5 x$10^{45}$ \\
3C 273/ T1    & $< 270 $  & 2.4 x$10^{46}$ & 5.7 x $10^{46}$ &  8 x $10^{46}$  &  90  & 30  & 0.8  & 1.9 x$10^{46}$ \\
\enddata

Column 1: Object  name and AGN type: type 1, 2 and LINER. \\
 Column 2: upper limit to the core size at 2 $\mu$m, set by the spatial resolution achieved in the K-band adaptive optics image. Only in Circinus  and NGC 1068  x is resolved at 2 $\mu$m, the quoted sizes are  from Prieto et al. 2004 and Weigelt et al. 2004 respectively.\\
 Column 3: IR luminosity  integrated over the high spatial resolution (HSR) data, from radio to the inflexion point  in the optical. It includes \textit{IRAS} and/or \textit{ISO} data only in the cases of NGC 3783, NGC 5506 and 3C 273 only (see text for details). \\
Column 4:  X-ray luminosity from the highest energy 
range  available in the literature. These are  as follows:\\
L(20 --100 keV) for Cen A, Circinus, NGC 1068 and NGC 5506  (Beckmann et al. 2006),  and for NGC 7582 and NGC 1566 (Landi et al. 2005); L(2 --10 keV) for NGC 1097 (Terashima et al. 2002); L(17--60 keV) for NGC 3783 and NGC 7469(Sazonov et al. 2007); L(20 -- 200 KeV) for 3C 273, average of two measurements at different epochs (Lichti et al. 1995). All the luminosities are normalised to $H_0 = 70$ km s$^{-1}$ Mpc$^{-1}$ when required  or to the adopted distance for the nearest objects, namelly Cen A, Circinus, NGC 1068, NGC 1097 and NGC 1566 (see object's section) . \\
 Column 5:  total luminosity defined as $L_{IR-HSR}  + L_{x( > 20 keV)}$.  Column 6:  optical-UV luminosity -blue bump- relative to $L_{IR-HSR}$. \\
 Column 7:  IR fraction of the total luminosity. In columns 8 and 9,   $L_{IRAS}$ is the  \textit{IRAS} luminosity following Sanders \& Mirabel's prescription (1996), and  effectively accounts for the  8 - 1000 $\mu$m region.
\end{deluxetable}

\begin{deluxetable}{lcccc}
\tabletypesize{\scriptsize}
\tablecaption{Extinction  towards the nucleus: comparative}
\label{Av}
\tablehead{\colhead{Name}   &
\colhead{$A_V$}  & 
\colhead{$A_V$} &  
\colhead{$A_V$} \\ 
 & J - K & 9.6 $\mu$m & N$_H$&   \\ 
}
\startdata

Type 2  \\
------------ \\
Cen A         &  5 & 14 & 62 \\ 
Circinus     & 6  & 21 & Compton-thick \\
NGC 1068   & 10-12 & 7 - 30 & Compton-thick \\
NGC 7582   &  9 & 20 &$>$70 \\
NGC 5506  &  5 & 15 & 17 \\

LINER \\
-----------\\
NGC 1097  & 1 & - & -\\

Type 1  \\
------------ \\
NGC 1566  & 7  &  - & -  \\
NGC 3783   & $<\sim0.5$ & - & - \\
NGC 7469  &  - & - & - \\
3C 273   & -  & -  & - \\
\enddata

$A_V$ derived from $J-K$ maps: nuclear extinction relative  to that in the galaxy  within the  central kpc region  (see text); $A_v$ from the optical depth at the Silicate 9.6 $\mu$m feature is derived from MIDI spectra in Cen A (Meisenheimer et al. 2007), Circinus (Tristram et al. 2007), and NGC 1068 (Raban et al. 2009); from  TIMMI2 spectra in   NGC 5506 and NGC 7582 (Siebenmorgen et al. 2004). In NGC 3783 no silicate feature in the interferometric MIDI spectrum is apparent (Beckert et al. 2008).  No report on the silicate optical depth was found for the remaining sources. The standard dust-to-gas ratio is applied  to infer the extinction from  the X-ray 
N(H). For those  with no entry, the Galaxy   N(H) is assumed in their X-ray fit. References to N(H) are  in the caption to table 1. 
\end{deluxetable}
\clearpage


\begin{deluxetable}{lccc}
\tabletypesize{\scriptsize}
\tablecaption{SED of Centaurus A core region}
\tablehead{\colhead{Origin}  & \colhead{Hz} & \colhead{Jy}  \\}
\startdata
     COMPTEL 5.1 MeV & 1.25$\times 10^{21}$ &  1.78$\times 10^{-8}$ \\
      COMPTEL 1.7 MeV & 3.98$\times 10^{20}$ &  1.0$\times 10^{-7}$ \\
     COMPTEL 0.86 MeV & 2.09$\times 10^{20}$ &  3.75$\times 10^{-7}$ \\
     INTEGRAL 20-100 keV & 1.45$\times 10^{19}$ &  2.4$\times 10^{-6}$ \\
     INTEGRAL 2-10 keV & 2.42$\times 10^{18}$ &  8.8$\times 10^{-6}$ \\
        CHANDRA 1 keV & 2.42$\times 10^{17}$ &  6.2$\times 10^{-5}$ \\
           WFPC2-F555W & 5.40$\times 10^{14}$ &  $<$ 6.4$\times 10^{-8}$ \\
           WFPC2-F814W & 3.68$\times 10^{14}$ &  7.0$\times 10^{-6}$ \\
            NACO-J & 2.30$\times 10^{14}$ &   0.0013 \\
            NACO-H & 1.76$\times 10^{14}$ &   0.0045 \\
            NACO-2.02  $\mu$m& 1.48$\times 10^{14}$ &    0.030 \\
            NACO-K & 1.36$\times 10^{14}$ &    0.031 \\
            NACO-L & 8.57$\times 10^{13}$ &     0.20 \\
           MIDI-8 $\mu$m & 3.70$\times 10^{13}$ &     0.60 \\
           MIDI-12 $\mu$m & 2.50$\times 10^{13}$ &      1.1 \\
           MIDI-13 $\mu$m & 2.30$\times 10^{13}$ &      1.3 \\
          MIDIcorre-8.3 $\mu$m & 3.60$\times 10^{13}$ &     0.47 \\
          MIDIcorre-9.3 $\mu$m & 3.20$\times 10^{13}$ &     0.28 \\
           MIDIcorre-10.4 $\mu$m & 2.88$\times 10^{13}$ &     0.25 \\
           MIDIcorre-11.4 $\mu$m & 2.63$\times 10^{13}$ &     0.43 \\
           MIDIcorre-12.6 $\mu$m & 2.30$\times 10^{13}$ &     0.62 \\
        VISIR-11.88 $\mu$m & 2.56$\times 10^{13}$ &      1.1 \\
        VISIR-18.72 $\mu$m & 1.70$\times 10^{13}$ &      2.3 \\
     VLBA-1997 & 2.22$\times 10^{10}$ &      1.9 \\
          VLBA-1997 & 8.40$\times 10^{9}$ &      1.7 \\
          VLBA-1999   & 8.4$\times 10^{9}$ &      2.32 \\
          VLBA-1999 & 5.0$\times 10^{9}$ &     0.83 \\
          VLBA-1999 & 2.2$\times 10^{9}$ &     1.03 \\
          ========== & & \\ 
               IRAS-12 $\mu$m & 2.50$\times 10^{13}$ &      22. \\
                IRAS-25 $\mu$m & 1.20$\times 10^{13}$ &      28. \\
                IRAS-60 $ \mu$m & 5.00$\times 10^{12}$ &  2.1$\times 10^{2}$ \\
                IRAS-100 $\mu$m & 3.00$\times 10^{12}$ &  4.1$\times 10^{2}$ \\
           ISO-170 $\mu$m & 1.76$\times 10^{12}$ &  5.4$\times 10^{2}$ \\
         SCUBA-350 $\mu$m & 8.66$\times 10^{11}$ &      7.7 \\
         SCUBA-450 $\mu$m & 6.67$\times 10^{11}$ &      7.9 \\
         SCUBA-750 $\mu$m & 4.07$\times 10^{11}$ &      8.1 \\
         SCUBA-850 $\mu$m & 3.50$\times 10^{11}$ &      8.1 \\
\enddata
 
Data sources and associated spatial resolutions are given in the object section. The large aperture data in the IR, shown with crosses in the SED in Fig.1, are added  at the end of the table.

\end{deluxetable}

\begin{deluxetable}{lccc}
\tabletypesize{\scriptsize}

\tablecaption{ SED of Circinus core region}
\tablehead{\colhead{Origin} &
\colhead{Hz} & \colhead{Jy}  \\}
\startdata

 INTEGRAL 40-100  keV & 1.69$\times 10^{19}$ &  5.5$\times 10^{-7}$ \\
   INTEGRAL 20-40 keV & 7.25$\times 10^{18}$ &  1.5$\times 10^{-6}$ \\
    INTEGRAL 2-10 keV & 1.45$\times 10^{18}$ &  8.9$\times 10^{-7}$ \\
        ROSAT 1.3 keV & 3.25$\times 10^{17}$ &  4.0$\times 10^{-6}$ \\
              NACO-J & 2.30$\times 10^{14}$ &      $<$ 0.0016 \\
              NACO-H & 1.76$\times 10^{14}$ &   0.0048 \\
              NACO-K & 1.36$\times 10^{14}$ &    0.019 \\
         NACO-2.42$\mu$m & 1.24$\times 10^{14}$ &    0.031 \\
              NACO-L & 7.90$\times 10^{13}$ &     0.38 \\
              NACO-M & 6.70$\times 10^{13}$ &      1.9 \\
        MIDIcorr-8 $\mu$m & 4.00$\times 10^{13}$ &     0.30 \\
      MIDIcorr-9.6 $\mu$m & 3.10$\times 10^{13}$ &     0.30 \\
       MIDIcorr-11 $\mu$m & 2.50$\times 10^{13}$ &     0.50 \\
       MIDIcorr-12 $\mu$m & 2.00$\times 10^{13}$ &     1.1 \\
            MIDI-8 $\mu$m & 3.70$\times 10^{13}$ &      6.0 \\
          MIDI-9.6 $\mu$m & 3.10$\times 10^{13}$ &      2.8 \\
       VISIR-11.88 $\mu$m & 2.50$\times 10^{13}$ &      9.3 \\
       VISIR-18.72 $\mu$m & 1.60$\times 10^{13}$ &      17.6 \\
            ATCA-3 cm & 1.00$\times 10^{10}$ &    0.050 \\
            ATCA-6 cm & 5.00$\times 10^{9}$ &    0.050 \\
           ATCA-13 cm & 2.30$\times 10^{9}$ &    0.070 \\
           ATCA-20 cm & 1.50$\times 10^{9}$ &     0.12 \\
           ========== & & \\ 

           IRAS-12 $\mu$m & 2.50$\times 10^{13}$ &      19. \\
           IRAS-25 $\mu$m & 1.20$\times 10^{13}$ &      68. \\
           IRAS-60 $\mu$m & 5.00$\times 10^{12}$ &  249 \\
          IRAS-100 $\mu$m & 3.00$\times 10^{12}$ &  316 \\
\enddata

Data sources and associated spatial resolutions are given in the object section. The large aperture data in the IR, shown with crosses in the SED in Fig.1, are added  at the end of the table.

\end{deluxetable}

\begin{deluxetable}{lccc}
\tabletypesize{\scriptsize}

\tablecaption{ SED of NGC 1068 core region}
\tablehead{\colhead{Origin} &
\colhead{Hz} & \colhead{Jy}  \\}
\startdata
 INTEGRAL-40-100 keV & 1.69$\times 10^{19}$ &  9.8$\times 10^{-8}$ \\
   INTEGRAL-20-40 keV & 7.25$\times 10^{18}$ &  1.3$\times 10^{-7}$ \\
      EXOSAT-2-10 keV & 1.45$\times 10^{18}$ &  3.7$\times 10^{-7}$ \\
        CHANDRA-1 keV & 2.42$\times 10^{17}$ &  1.5$\times 10^{-7}$ \\
                   NACO-J & 2.30$\times 10^{14}$ &   $<$0.0024 \\
              NACO-H & 1.76$\times 10^{14}$ &   0.0056 \\
              NACO-K & 1.36$\times 10^{14}$ &    0.056 \\
           NACO-2.42 & 1.24$\times 10^{14}$ &     0.12 \\
           NACO-M & 6.70$\times 10^{13}$ &      1.5 \\
      MIDIcorr-8 $\mu$m & 3.75$\times 10^{13}$ &     1.9\\
      MIDIcorr-9.6 $\mu$m & 3.1$\times 10^{13}$ &     0.4 \\
       MIDIcorr-12 $\mu$m & 2.5$\times 10^{13}$ &     1.2 \\
       MIDIcorr-13 $\mu$m & 2.3$\times 10^{13}$ &     1.9 \\
            Subaru-8.72$\mu$m & 3.4$\times 10^{13}$ &      5.81 \\
            Subaru-9.69$\mu$m & 3.1$\times 10^{13}$ &      4.03 \\          
            Subaru-10.38$\mu$m & 2.9$\times 10^{13}$ &    3.82 \\
             Subaru-11.66$\mu$m & 2.6$\times 10^{13}$ &     10.2 \\
            Subaru-12.33$\mu$m & 2.4$\times 10^{13}$ &      9.3 \\
                   Subaru-18.5 $\mu$m & 1.60$\times 10^{13}$ &      9.6 \\
           Keck-25 $\mu$m & 1.20$\times 10^{13}$ &      9.6 \\
        IRAM-1 mm & 2.30$\times 10^{11}$ &    0.022 \\
          IRAM-3 mm & 1.15$\times 10^{11}$ &    0.036 \\
               VLA-A & 4.30$\times 10^{10}$ &    0.013 \\
               VLA-A & 2.25$\times 10^{10}$ &    0.016 \\
                VLBA & 8.40$\times 10^{9}$ &   0.0054 \\
                VLBA & 5.00$\times 10^{9}$ &   0.0091 \\
              VLBA & 1.70$\times 10^{9}$ &   $<$0.0026 \\
          ========== & & \\ 

          IRAS-12 $\mu$m & 2.50$\times 10^{13}$ &      40. \\
           IRAS-25 $\mu$m & 1.20$\times 10^{13}$ &      85. \\
             IRAS-60 $\mu$m& 5.00$\times 10^{12}$ &  1.8$\times 10^{2}$ \\
          IRAS-100 $\mu$m & 3.00$\times 10^{12}$ &  2.8$\times 10^{2}$ \\
           ISO-170 $\mu$m & 1.76$\times 10^{12}$ &  3.9$\times 10^{2}$ \\
           MKO-390 $\mu$m & 7.69$\times 10^{11}$ &      30. \\
          CTIO-540 $\mu$m & 5.55$\times 10^{11}$ &      7.0 \\

\enddata
 
  Data sources and  spatial scales are given in the object section.
The large aperture data in the IR, shown with crosses in the SED in Fig.1, are added  at the end of the table.

\end{deluxetable}

\begin{deluxetable}{lccc}
\tabletypesize{\scriptsize}

\tablecaption{ SED of NGC 1097 core region}
\tablehead{\colhead{Origin} &
\colhead{Hz} & \colhead{Jy}  \\}
\startdata
       ASCA-2-10 keV & 2.42$\times 10^{18}$ &  8.5$\times 10^{-8}$ \\
       ASCA-0.5-4 keV & 2.42$\times 10^{17}$ &  2.4$\times 10^{-7}$ \\
         WFPC2-F218W & 1.40$\times 10^{15}$ &  1.8$\times 10^{-5}$ \\
          WFPC-F555W & 5.41$\times 10^{14}$ &  8.8$\times 10^{-5}$ \\
         ACSWF-F814W & 3.60$\times 10^{14}$ &  0.00074 \\
              NACO-J & 2.30$\times 10^{14}$ &   0.0011 \\
              NACO-H & 1.76$\times 10^{14}$ &   0.0027 \\
              NACO-K & 1.36$\times 10^{14}$ &   0.0039 \\
              NACO-L & 7.90$\times 10^{13}$ &    0.011 \\
         VISIR-11.88  $\mu$m & 2.50$\times 10^{13}$ &    0.025 \\
         VISIR-18.72  $\mu$m & 1.60$\times 10^{13}$ &    0.041 \\
           VLA-B & 1.50$\times 10^{10}$ &   0.0056 \\
          VLA-A& 8.40$\times 10^{9}$ &   0.0031 \\
          VLA-A & 4.80$\times 10^{9}$ &   0.004 \\
          VLA-B & 1.40$\times 10^{9}$ &   0.0012 \\
          ========== & & \\ 
             IRAS-12  $\mu$m & 2.50$\times 10^{13}$ &      3.0 \\
             IRAS-25  $\mu$m & 1.20$\times 10^{13}$ &      7.3 \\
             IRAS-60  $\mu$m & 5.00$\times 10^{12}$ &      53. \\
            IRAS-100  $\mu$m & 3.00$\times 10^{12}$ &  1.0$\times 10^{2}$ \\
         Spitzer-160  $\mu$m & 1.92$\times 10^{12}$ &  1.5$\times 10^{2}$ \\
           SCUBA-850  $\mu$m & 3.53$\times 10^{11}$ &      1.4 \\
\enddata

 Data sources and  spatial scales are given in the object section. The large aperture data in the IR, shown with crosses in the SED in Fig.1, are added  at the end of the table.

\end{deluxetable}

\begin{deluxetable}{lccc}
\tabletypesize{\scriptsize}

\tablecaption{ SED of NGC 7582 core region}
\tablehead{\colhead{Origin} &
\colhead{Hz} & \colhead{Jy}  \\}
\startdata
 SAX 20-100  keV & 1.45$\times 10^{19}$ &  8.1$\times 10^{-7}$ \\
         XMM 2-12 keV & 1.69$\times 10^{18}$ &  1.4$\times 10^{-6}$ \\
ASCA 0.5-2 keV   3.03$\times 10^{17}$	&	1.46$\times 10^{-7}$ \\
         WFPC2-F606W & 5.00$\times 10^{14}$ &  3.0$\times 10^{-6}$ \\
        NICMOS-F160W & 1.88$\times 10^{14}$ &    0.011 \\
         NACO-2.06 $\mu$m & 1.46$\times 10^{14}$ &    0.018 \\
         NACO-L & 7.89$\times 10^{13}$ &    0.096 \\
         NACO-4.05 $\mu$m & 7.41$\times 10^{13}$ &     0.11 \\
   TIMMI2-spec-8.5 $\mu$m & 3.53$\times 10^{13}$ &     0.3 \\
     TIMMI2-spec-9 $\mu$m & 3.33$\times 10^{13}$ &     0.2 \\
   TIMMI2-spec-9.6 $\mu$m & 3.13$\times 10^{13}$ &    0.08 \\
    TIMMI2-spec-11 $\mu$m & 2.73$\times 10^{13}$ &     0.2 \\
TIMMI2-spec-12 $\mu$m & 2.50$\times 10^{13}$ &     0.4 \\
         VISIR-11.88 $\mu$m & 2.52$\times 10^{13}$ &     0.40 \\
         VISIR-18.72 $\mu$m & 1.60$\times 10^{13}$ &     0.55 \\
        VLA-A & 8.40$\times 10^{9}$ &   0.0069 \\
           VLA-A & 5.00$\times 10^{9}$ &   0.0095 \\
          ========== & & \\ 
           IRAS-12  $\mu$m & 2.50$\times 10^{13}$ &      2.3 \\
           IRAS-25 $\mu$m & 1.20$\times 10^{13}$ &      7.4 \\
           IRAS-60 $\mu$m & 5.00$\times 10^{12}$ &      5.2 \\
          IRAS-100 $\mu$m & 3.00$\times 10^{12}$ &      8.3 \\
\enddata
 
 Data sources and  spatial scales are given in the object section.
The large aperture data in the IR, shown with crosses in the SED in Fig.1, are added  at the end of the table.

\end{deluxetable}

\begin{deluxetable}{lccc}
\tabletypesize{\scriptsize}

\tablecaption{ SED of NGC  5506         core region}
\tablehead{\colhead{Origin} &
\colhead{Hz} & \colhead{Jy}  \\}
\startdata
  INTEGRAL-40-100 keV & 1.69$\times 10^{19}$ &  2.2$\times 10^{-7}$ \\
       INTEGRAL- 20-40 keV & 7.25$\times 10^{18}$ &  5.8$\times 10^{-7}$ \\
          INTEGRAL-2-10 keV & 1.45$\times 10^{18}$ &  6.1$\times 10^{-6}$ \\
        Einstein-0.2-4 keV & 5.25$\times 10^{17}$ &  2.1$\times 10^{-6}$ \\
           WFPC2-V+R & 5.00$\times 10^{14}$ &  $<$ 2.9$\times 10^{-5}$ \\
            NACO-J & 2.30$\times 10^{14}$ &    0.013 \\
              NACO-H & 1.76$\times 10^{14}$ &    0.053 \\
              NACO-K & 1.36$\times 10^{14}$ &    0.080 \\
              NACO-L & 7.90$\times 10^{13}$ &     0.29 \\
      ISO-spec-6 $\mu$m & 5.00$\times 10^{13}$ &     0.70 \\
      TIMMI2-spec-9.6 $\mu$m & 3.0$\times 10^{13}$    &     0.5\\
         VISIR-11.88 $\mu$m & 2.50$\times 10^{13}$ &     0.9 \\
      VISIR-18.72 $\mu$m & 1.70$\times 10^{13}$ &      1.4 \\
          VLBA-1997 & 8.33$\times 10^{9}$ &    0.022 \\
            VLBI-1994 & 5.00$\times 10^{9}$ &    0.034 \\
            VLBA-1997 & 5.00$\times 10^{9}$ &    0.030 \\
            VLBA-2000 & 5.00$\times 10^{9}$ &    0.042 \\
         PTI-1990 & 2.30$\times 10^{9}$ &    0.087 \\
           VLBA-1997 & 1.60$\times 10^{9}$ &    0.024 \\
        VLBA-2000 & 1.60$\times 10^{9}$ &    0.046 \\
          ========== & & \\ 
              IRAS-12 $\mu$m & 2.50$\times 10^{13}$ &      1.3 \\
              IRAS-25 $\mu$m & 1.20$\times 10^{13}$ &      4.2 \\
              IRAS-60 $\mu$m & 5.00$\times 10^{12}$ &      8.4 \\
             IRAS-100 $\mu$m & 3.00$\times 10^{12}$ &      8.9 \\

\enddata

 Data sources and corresponding spatial scales are given in the object section.
The large aperture data in the IR, shown with crosses in the SED in Fig.1, are added  at the end of the table.

\end{deluxetable}

\begin{deluxetable}{lccc}
\tabletypesize{\scriptsize}

\tablecaption{ SED of NGC 1566 core region}
\tablehead{\colhead{Origin} &
\colhead{Hz} & \colhead{Jy}  \\}
\startdata
   
   SAX-20-100 keV & 1.45$\times 10^{19}$ &  5.6$\times 10^{-7}$ \\
     Einstein-2-4 keV & 5.25$\times 10^{17}$ &  2.8$\times 10^{-6}$ \\
           FOS-1300 A & 2.30$\times 10^{15}$ &  5.6$\times 10^{-5}$ \\
           FOS-1400 A & 2.10$\times 10^{15}$ &  8.0$\times 10^{-5}$ \\
           FOS-1800 A & 1.67$\times 10^{15}$ &  0.00016 \\
           FOS-2100 A & 1.40$\times 10^{15}$ &  0.00021 \\
         WFPC2-160BW & 1.87$\times 10^{15}$ &  0.00011 \\
         WFPC2-F336W & 8.92$\times 10^{14}$ &  0.00038 \\
         WFPC2-F547M & 5.48$\times 10^{14}$ &  0.00050 \\
         WFPC2-F555W & 5.40$\times 10^{14}$ &  0.00053 \\
         WFPC2-F814W & 3.68$\times 10^{14}$ &  0.00079 \\
              NACO-J & 2.37$\times 10^{14}$ &   0.0011 \\
              NACO-K & 1.38$\times 10^{14}$ &   0.0021 \\
              NACO-L & 7.90$\times 10^{13}$ &   0.0078 \\
       VISIR-11.88 $\mu$m & 2.50$\times 10^{13}$ &    0.059 \\
       VISIR-18.72 $\mu$m & 1.60$\times 10^{13}$ &     0.11 \\
         ATCA-8.4 GHz & 8.40$\times 10^{9}$ &   0.0080 \\
         PTI-2.3 GHz & 2.30$\times 10^{9}$ &   0.0050 \\
         PTI-1.7 GHz & 1.70$\times 10^{9}$ &   $<$0.006 \\
         ========== & & \\
                IRAS-12 $\mu$m & 2.50$\times 10^{13}$ &      1.9 \\
                IRAS-25 $\mu$m & 1.20$\times 10^{13}$ &      3.0 \\
                IRAS-60 $\mu$m & 5.00$\times 10^{12}$ &      22. \\
                IRAS-100 $\mu$m & 3.00$\times 10^{12}$ &      58. \\
    Spitzer-160 $\mu$m & 1.92$\times 10^{12}$ &  1.0$\times 10^{2}$ \\
 \enddata

Data sources and  spatial scales are given in the object section.
The large aperture data in the IR, shown with crosses in the SED in Fig.1, are added  at the end of the table.

\end{deluxetable}

\begin{deluxetable}{lccc}
\tabletypesize{\scriptsize}
\tablecaption{ SED of NGC 3783 core region}
\tablehead{\colhead{Origin} &
\colhead{Hz} & \colhead{Jy}  \\}
\startdata
      OSSE 50-150 keV & 2.42$\times 10^{19}$ &  2.9$\times 10^{-7}$ \\
   INTEGRAL 17-60 keV & 9.31$\times 10^{18}$ &  1.3$\times 10^{-6}$ \\
         XMM 2-10 keV & 1.45$\times 10^{18}$ &  3.9$\times 10^{-6}$ \\
   Einstein 0.2-4 Kev & 5.25$\times 10^{17}$ &  7.9$\times 10^{-6}$ \\
          FUSE-1040 A & 3.00$\times 10^{15}$ &   0.001 \\
          STIS-1150 A & 2.70$\times 10^{15}$ &   0.0016 \\
          STIS-1346 A & 2.20$\times 10^{15}$ &   0.0023 \\
          STIS-1480 A & 2.00$\times 10^{15}$ &   0.0025 \\
          STIS-2279 A & 1.30$\times 10^{15}$ &   0.0045 \\
          STIS-2419 A & 1.20$\times 10^{15}$ &   0.0056 \\
          STIS-2640 A & 1.13$\times 10^{15}$ &   0.0065 \\
          STIS-2900 A & 1.00$\times 10^{15}$ &   0.0075 \\
          STIS-3100 A & 9.70$\times 10^{14}$ &   0.0076 \\
          LCO-U & 8.19$\times 10^{14}$ &   0.0056 \\
           ACS-F547M & 5.48$\times 10^{14}$ &   0.0060 \\
           ACS-F550M & 5.45$\times 10^{14}$ &   0.0066 \\
         WFPC2-F814W & 3.68$\times 10^{14}$ &   0.0099 \\
         NACO-J & 2.30$\times 10^{14}$ &    0.023 \\
         NACO-K & 1.36$\times 10^{14}$ &    0.073 \\
         NACO-L & 7.90$\times 10^{13}$ &     0.17 \\
       VISIR-11.88 $\mu$m & 2.50$\times 10^{13}$ &     0.54 \\
       VISIR-18.72 $\mu$m & 1.60$\times 10^{13}$ &      1.47 \\
       VLA-A & 8.40$\times 10^{9}$ &   0.0080 \\
          VLA-A & 4.80$\times 10^{9}$ &    0.013 \\
           VLA-A  & 1.40$\times 10^{9}$ &    0.023 \\
          ========== & & \\ 
           IRAS-12 $\mu$m & 2.50$\times 10^{13}$ &     0.84 \\
           IRAS-25 $\mu$m & 1.20$\times 10^{13}$ &      2.5 \\
           IRAS-60 $\mu$m & 5.00$\times 10^{12}$ &      3.3 \\
          IRAS-100 $\mu$m & 3.00$\times 10^{12}$ &      4.9 \\
\enddata
 
 Data sources and corresponding spatial scales are given in the object section.
The large aperture data in the IR, shown with crosses in the SED in Fig.1, are added  at the end of the table.
\end{deluxetable}

\begin{deluxetable}{lccc}
\tabletypesize{\scriptsize}
\tablecaption{ SED of NGC 7469 core region}
\tablehead{\colhead{Origin} &
\colhead{Hz} & \colhead{Jy}  \\}
\startdata

   INTEGRAL-17-60 keV & 9.31$\times 10^{18}$ &  5.1$\times 10^{-7}$ \\
   XMM-2-10 keV & 1.45$\times 10^{18}$ &  2.9$\times 10^{-6}$ \\
   ROSAT-0.1-2.4 keV & 2.42$\times 10^{17}$ &  9.1$\times 10^{-6}$ \\
                  WFPC2-F218W & 1.36$\times 10^{15}$ &   0.0027 \\
                   ACS-F330W & 8.92$\times 10^{14}$ &   0.0031 \\
                   WFPC2-F547M & 5.47$\times 10^{14}$ &   0.0029 \\
                   ACS-F550M & 5.38$\times 10^{14}$ &   0.0022 \\
                    ACS-F814W & 3.75$\times 10^{14}$ &   0.0042 \\
                 NACO-J & 2.37$\times 10^{14}$ &   0.0080 \\
                  NACO-H & 1.81$\times 10^{14}$ &    0.015 \\
                  NICMOS-F187N & 1.60$\times 10^{14}$ &    0.019 \\
                 NACO-K & 1.38$\times 10^{14}$ &    0.020 \\
                 NACO-L & 7.89$\times 10^{13}$ &    0.084 \\
                NACO-4.05 $\mu$m & 7.41$\times 10^{13}$ &    0.096 \\
                VISIR-11.88 $\mu$m & 2.53$\times 10^{13}$ &     0.53 \\
               VISIR-18.72 $\mu$m & 1.60$\times 10^{13}$ &      1.27 \\
                VLA-A & 1.49$\times 10^{10}$ &    0.011 \\
                VLA-A & 8.40$\times 10^{9}$ &    0.015 \\
                     MERLIN & 5.00$\times 10^{9}$ &    0.012 \\
                     VLBI & 2.30$\times 10^{9}$ &    0.014 \\
                  VLBI & 1.70$\times 10^{9}$ &    0.021 \\
         ========== & & \\ 
                  IRAS-12 $\mu$m & 2.50$\times 10^{13}$ &      1.6 \\
                  Spitzer-15 $\mu$m & 2.00$\times 10^{13}$ &      1.5 \\
                  Spitzer-20 $\mu$m & 1.50$\times 10^{13}$ &      3.2 \\
                     IRAS-25 $\mu$m& 1.20$\times 10^{13}$ &      6.0 \\
                  Spitzer-30 $\mu$m& 9.99$\times 10^{12}$ &      7.8 \\
                     IRAS-60 $\mu$m & 5.00$\times 10^{12}$ &      27.3 \\
                     IRAS-100 $\mu$m & 3.00$\times 10^{12}$ &      35.2 \\
                   Caltech-350  $\mu$m & 8.57$\times 10^{11}$ &      2.23 \\
                  SCUBA-850  $\mu$m & 3.53$\times 10^{11}$ &     0.19 \\

\enddata

 Data sources and  spatial scales are given in the object section.   The large aperture data in the IR, shown with crosses in the SED in Fig.1, are added  at the end of the table.

\end{deluxetable}

\begin{deluxetable}{lccc}
\tabletypesize{\scriptsize}
\tablecaption{ SED of 3C 273 core region}
\tablehead{\colhead{Origin} &
\colhead{Hz} & \colhead{Jy}  \\}
\startdata
                 1 GeV & 2.42$\times 10^{23}$ &  1.4$\times 10^{-11}$ \\
              300 MeV & 7.25$\times 10^{22}$ &  8.8$\times 10^{-11}$ \\
              100 MeV & 2.42$\times 10^{22}$ &  4.8$\times 10^{-10}$ \\
                3 MeV & 7.25$\times 10^{20}$ &  5.6$\times 10^{-8}$ \\
                1 MeV & 2.40$\times 10^{20}$ &  1.8$\times 10^{-7}$ \\
              500 keV & 1.21$\times 10^{20}$ &  2.9$\times 10^{-7}$ \\
              100 keV & 2.42$\times 10^{19}$ &  1.1$\times 10^{-6}$ \\
               50 keV & 1.21$\times 10^{19}$ &  1.6$\times 10^{-6}$ \\
               20 keV & 4.84$\times 10^{18}$ &  3.3$\times 10^{-6}$ \\
              10 keV & 2.42$\times 10^{18}$ &  4.2$\times 10^{-6}$ \\
                5 keV & 1.21$\times 10^{18}$ &  5.9$\times 10^{-6}$ \\
                2 keV & 4.84$\times 10^{17}$ &  8.7$\times 10^{-6}$ \\
                1 keV & 2.42$\times 10^{17}$ &  1.4$\times 10^{-5}$ \\
              0.5 keV & 1.21$\times 10^{17}$ &  2.9$\times 10^{-5}$ \\
              0.2 keV & 4.84$\times 10^{16}$ &  9.5$\times 10^{-5}$ \\
           0.1 keV & 2.42$\times 10^{16}$ &  2.48$\times 10^{-4}$ \\
            1300 A & 2.31$\times 10^{15}$ &    0.012 \\
               2100 A & 1.43$\times 10^{15}$ &    0.019 \\
               3000 A & 9.99$\times 10^{14}$ &    0.025 \\
                  U & 8.57$\times 10^{14}$ &    0.027 \\
                   B & 7.00$\times 10^{14}$ &    0.027 \\
                   V & 5.50$\times 10^{14}$ &    0.029 \\
                   R & 4.43$\times 10^{14}$ &    0.027 \\
                   I & 3.37$\times 10^{14}$ &    0.028 \\
              NACO-J & 2.30$\times 10^{14}$ &    0.032 \\
              NACO-H & 1.76$\times 10^{14}$ &    0.047 \\
              NACO-K & 1.36$\times 10^{14}$ &    0.068 \\
              NACO-L & 7.90$\times 10^{13}$ &     0.17 \\
              NACO-M & 6.70$\times 10^{13}$ &     0.17 \\ 
         VLBI-2 mm& 1.47$\times 10^{11}$ &      2.2 \\
          VLBI-3 mm & 1.00$\times 10^{11}$ &      8.0 \\
           VLBI & 8.60$\times 10^{10}$ &      7.88 \\
          VLBI & 4.30$\times 10^{10}$ &      6.66\\
          VLBI & 2.20$\times 10^{10}$ &      4.99 \\
          VLBA & 1.50$\times 10^{10}$ &      9.18 \\
        VLBI & 5.00$\times 10^{9}$ &      12.7 \\
          ========== & & \\ 
       ISO-6.7 $\mu$m & 4.44$\times 10^{13}$ &     0.19 \\
      ISO-14.3 $\mu$m & 2.00$\times 10^{13}$ &     0.29 \\
        ISO-80 $\mu$m & 3.70$\times 10^{12}$ &      1.29 \\
       ISO-100 $\mu$m & 2.90$\times 10^{12}$ &      1.35 \\
       ISO-120 $\mu$m & 2.52$\times 10^{12}$ &      1.55 \\
       ISO-150 $\mu$m & 1.86$\times 10^{12}$ &      1.11 \\
       ISO-170 $\mu$m & 1.72$\times 10^{12}$ &      1.29 \\
       ISO-180 $\mu$m & 1.62$\times 10^{12}$ &      1.06 \\
       ISO-200 $\mu$m & 1.47$\times 10^{12}$ &      1.09 \\
 \enddata

   The SED from 1 GeV to the I-band is taken directly from  Turler et al. (1999). Further data are compiled in this work and the  sources  are given in the corresponding section for this object.  The large aperture data in the IR, shown with crosses in the SED in Fig.1, are added  at the end of the table.

\end{deluxetable}

\begin{deluxetable}{lccc}
\tabletypesize{\scriptsize}
\tablecaption{Galaxies observed with adaptive optics in the near-IR with VLT / NACO}
\label{observations}
\tablehead{\colhead{Name} & 
\colhead{Filters (observing date) } &
\colhead{Reference} \\}

\startdata
Cen A      &  J, H, NB-2.02 $\mu$m, L (March - May  2003), K (March  2004) & Haering-Neumayer et al. 2006\\
Circinus   & J, K, NB-2.42 $\mu$m, L,M (March - May 2003) & Prieto et al. 2004\\
NGC 1068 &  K, M (Nov -Dec 2002), J, H (Jan 2004)   J, NB-2.42 $\mu$m  (Jan 2005) & Hoenig et al.  2008 \\ 
NGC 7582 & 2.06 $\mu$m (May - Jun 2005), L, NB-4.05 $\mu$m (Dec 2005) & Fernandez-Ontiveros et al.  2009 \\
NGC 5506 &  J, K, L, M (Jun 2003) & this work\\
NGC 1097 & J, H, K ( Aug 2002), L (Jan 2005) & Prieto et al. 2005\\
NGC 1566 & K, L (Jan 2005), J( Nov 2005) & this work \\
NGC 3783 & J,K,L (Jan 2005) & this work \\
NGC 7469 & J, H, K (Nov 2002), L, NB-4.05 $\mu$m (Dec 2005 )  & this work \\
3C 273       & J,K,L, M  (May 2003) & this work \\
\enddata

Galaxies sorted by distance.
\end{deluxetable}

\end{document}